\documentclass[prc,preprint, showpacs,amsmath,amssymb, floatfix]{revtex4}

\usepackage{graphicx}

\def\beq{\begin{equation}}
\def\eeq{\end{equation}}
\def\ba{\begin{eqnarray}}
\def\ea{\end{eqnarray}}

\def\v8p{v_8^\prime}

\def\sigij{\boldsigma_i \cdot \boldsigma_j}
\def\tauij{\boldtau_i \cdot \boldtau_j}
\def\chits{\chi_{\tau}} 
\def\chipu{\chi_p} 
\def\chinu{\chi_n} 
\def\gfa{f^c}
\def\gfb{f^{\tau}}
\def\gfc{f^{\sigma}}
\def\gfd{f^{\sigma \tau}}
\def\gfe{f^t}
\def\gff{f^{t \tau}}
\def\gfooadc{F^{0 0,\sigma}_{d,a}}
\def\gfoTadc{F^{0 1,\sigma}_{d,a}}
\def\gfToadc{F^{1 0,\sigma}_{d,a}}
\def\gfTTadc{F^{1 1,\sigma}_{d,a}}
\def\gfooadt{F^{0 0,A}_{d,a}}
\def\gfoTadt{F^{0 1,A}_{d,a}}
\def\gfToadt{F^{1 0,A}_{d,a}}
\def\gfTTadt{F^{1 1,A}_{d,a}}
\def\gfoojdc{F^{0 0,\sigma}_{d,j}}
\def\gfoTjdc{F^{0 1,\sigma}_{d,j}}
\def\gfTojdc{F^{1 0,\sigma}_{d,j}}
\def\gfTTjdc{F^{1 1,\sigma}_{d,j}}
\def\gfoojdt{F^{0 0,A}_{d,j}}
\def\gfoTjdt{F^{0 1,A}_{d,j}}
\def\gfTojdt{F^{1 0,A}_{d,j}}
\def\gfTTjdt{F^{1 1,A}_{d,j}}
\def\gfooaec{F^{0 0,\sigma}_{e,a}}
\def\gfoTaec{F^{0 1,\sigma}_{e,a}}
\def\gfToaec{F^{1 0,\sigma}_{e,a}}
\def\gfTTaec{F^{1 1,\sigma}_{e,a}}
\def\gfooaet{F^{0 0,A}_{e,a}}
\def\gfoTaet{F^{0 1,A}_{e,a}}
\def\gfToaet{F^{1 0,A}_{e,a}}
\def\gfTTaet{F^{1 1,A}_{e,a}}
\def\gfoojec{F^{0 0,\sigma}_{e,j}}
\def\gfoTjec{F^{0 1,\sigma}_{e,j}}
\def\gfTojec{F^{1 0,\sigma}_{e,j}}
\def\gfTTjec{F^{1 1,\sigma}_{e,j}}
\def\gfoojet{F^{0 0,A}_{e,j}}
\def\gfoTjet{F^{0 1,A}_{e,j}}
\def\gfTojet{F^{1 0,A}_{e,j}}
\def\gfTTjet{F^{1 1,A}_{e,j}}
\def\sigab{\langle \boldsigma_a \rangle}
\def\siga{\langle \widetilde{\boldsigma_a} \rangle}
\def\Atb{\langle \mathbf{A}_t \rangle}
\def\At{\langle \widetilde{\mathbf{A}_t} \rangle}
\def\PhiI{|\Phi_I \rangle }
\def\PhiF{|\Phi_F \rangle }
\def\gtav{\eta_{GT}}
\def\tauaz{\langle \tau_a^z \rangle}
\def\av{\eta_{NA}}

\newcommand{\boldsigma}{\mbox{\boldmath$\sigma$}}
\newcommand{\boldtau}{\mbox{\boldmath$\tau$}}

\newcommand{\boldnabla}{\mbox{\boldmath$\nabla$}}

\begin{document}

\title{Quenching of Weak Interactions in Nucleon Matter}
\author {S. Cowell and V. R. Pandharipande } 
\affiliation{ Department of Physics,  
  University of Illinois at Urbana-Champaign, \\
        1110 W. Green St., Urbana, IL 61801, U.S.A.}

\date{\today}

\begin{abstract}

We have calculated the one-body Fermi and Gamow-Teller charge-current, and vector and 
axial-vector neutral-current nuclear matrix elements in nucleon matter at 
densities of 0.08, 0.16 and 0.24 fm$^{-3}$ and proton fractions ranging from 0.2  to
0.5.  The correlated states for nucleon matter are obtained  by operating on
Fermi-gas states by a symmetrized product of pair  correlation operators determined from
variational calculations with the Argonne v18 and Urbana IX two-
and three-nucleon interactions.  The squares of the charge current matrix elements 
are found to be quenched by 20 to 25 \% by the
short-range correlations in nucleon matter.  Most of the quenching is due to
spin-isospin  correlations induced by the pion exchange interactions which  change
the isospins and spins of the nucleons.  A large part of it can be  related to the
probability for a spin up proton quasi-particle to be a  bare spin up/down
proton/neutron.  Within the interval considered  the 
charge current matrix elements do not have significant dependence  on the matter
density, proton fraction and magnitudes of nucleon momenta; however, they  do
depend upon momentum transfer.  The neutral current matrix elements have a 
significant dependence on the proton fraction.  
We also calculate the matrix elements of the 
nuclear Hamiltonian  in the same correlated basis.  These provide relatively mild 
effective interactions which give the variational energies in the Hartree-Fock
approximation.  The calculated two-nucleon effective
interaction describes the spin-isospin  susceptibilities of nuclear and neutron
matter fairly accurately.  However $\geq$ 3-body terms are necessary to reproduce the
compressibility.   Realistic calculations of weak
interaction rates in nucleon matter  can presumably be carried out using 
the effective  operators and interactions studied here. 
All presented results use the  simple 2-body cluster approximation to calculate
the correlated basis matrix elements.  This allows for a clear discussion of
the physical effects in the effective  operators and interactions. 
\end{abstract}
\pacs{21.30.Fe, 23.40.Hc, 26.50.+x}
\maketitle

\section{Introduction}
 
Weak interactions in nucleon matter occur during the beta-decay of nuclei, 
electron and muon capture reactions, neutrino-nucleus scattering 
and in various astrophysical environments, such as evolving stars, 
neutron stars and supernovae.
They have been studied since Fermi proposed the first theory of beta-decay
in 1934.  Recently there has been much interest in weak interactions 
in the sun \cite{sun1,sun2}, those  
of $^{12}$C and $^{16}$O due to their use in neutrino detectors searching for neutrino 
oscillations \cite{KARMEN,LSND1,LSND2,K2K}, and in interactions of neutrinos with dense 
matter in neutron stars and supernovae \cite{PLSV01}.  Low energy weak 
interactions proceed mainly via the nuclear matrix elements of the following 
four one-body operators: 
\ba
O_F = \sum_i O_F(i) &=& \sum_i \boldtau^{\pm}_i e^{i{\bf q}\cdot {\bf r}_i}~, \\
\mathbf{O}_{GT} = g_A \sum_i \mathbf{O}_{GT} (i) &=& g_A \sum_i \boldtau^{\pm}_i \boldsigma_i 
e^{i{\bf q}\cdot {\bf r}_i}~, \\
O_{NV} = \sum_i O_{NV}(i) &=& \sum_i \left( - sin^2 \theta_W + \frac{1}{2} 
(1-2 sin^2 \theta_W)\tau_{i}^z \right) 
e^{i{\bf q}\cdot {\bf r}_i}~, \\
\mathbf{O}_{NA} = g_A \sum_i \mathbf{O}_{NA}(i) &=& g_A \sum_i \frac{1}{2} \tau_{i}^z \boldsigma_i 
e^{i{\bf q}\cdot {\bf r}_i}~. 
\ea
Here $i$ is the nucleon number label and ${\bf q}$ is the momentum given by
the  weak boson to the nucleon.   The Fermi coupling constant multiplying these
operators is omitted for brevity,  $\theta_W$ is the electroweak mixing angle, 
and $g_A$ is the ratio of the weak axial vector and Fermi coupling constants of
the nucleon.  The four operators are called Fermi (F), Gamow-Teller (GT), 
neutral-vector (NV) and neutral-axial-vector (NA).  In the nonrelativistic
domain,  neglecting weak pair currents, the interaction of low energy neutrinos
with nuclei and nucleon  matter and nuclear beta-decay rates are proportional
to the square of the matrix elements of these operators between initial and
final nuclear states. 

Due to the strong forces, nuclear wave functions are highly correlated 
\cite{FPPWSA96,AP97}, and it is difficult to calculate the nuclear matrix 
elements.  Using quantum Monte Carlo and Faddeev methods to calculate nuclear 
wave functions from realistic models of nuclear forces, the beta-decay matrix elements 
have been calculated for light nuclei with $A \leq 7$ \cite{ppcap,SW02}.  The 
calculated values for $^3$H, $^6$Li and $^7$Be are within  
5 \% of the observed, and better agreement is obtained after including 
weak pair currents.  The weak muon capture by $^3$He has also been calculated \cite{mucap} 
with realistic wave functions with similar success. 

However, complete many-body calculations are not yet possible for nuclei like 
$^{12}$C and heavier, as well as for nucleon matter.  Most studies of 
weak interactions in these systems use effective interactions and 
shell-model and Fermi-gas wave functions in finite nuclei and nucleon matter respectively.   
The random phase approximation (RPA) is commonly used.  The pioneering work on 
GT transitions has been reviewed by Arima {\em et. al} \cite{Arima87}.  Some of the 
recent works are: \cite{KLV99,HT00} in $^{12}$C, \cite{LDRAK95,MPCZ96} in 
$pf$-shell, and \cite{RPL99} in neutron stars and supernovae.  Typically 
the calculated rate of weak interactions is larger than observed; for 
example, a factor of $\sim$ 0.6 brings the calculated 
$pf$-shell GT transition rates in agreement with experiment.  
Recent LSND results of charged current reaction cross sections of 
$\nu_e$ \cite{LSND1} and $\nu_{\mu}$ \cite{LSND2} on $^{12}$C are lower than the  
theoretical expectations by up to 20 \%. 

This is not surprising since effective operators which take into account the 
effects of short range correlations, and not the bare operators given by 
Eq. (1) to (4), must be used along with effective interactions as is well known 
from the works of Arima and collaborators \cite{Arima87}.
In nuclei near the line of 
stability the observed spectra and beta-decay rates have been used to 
model the effective interactions and operators, but for neutron stars and 
supernovae matter we have to calculate them from realistic models of 
nuclear forces.  In $pf$-shell and heavier nuclei the effective interaction 
is also obtained from bare forces \cite{KB}. 

There are several ways to obtain consistent sets of effective operators and 
interactions starting from a bare nuclear Hamiltonian.  For example, one can 
introduce a model space and employ the Lee-Suzuki similarity transformation
\cite{LeeS} as in the no core shell model type approach \cite{NCSM}.   
In this theory the effective operators and interactions take into account the 
truncated Hilbert space.  They are used in the retained model space to predict 
the observables.  In the present work we use the correlated basis (CB) approach 
\cite{FP87,FP88}, evolved out of variational theories of quantum liquids 
\cite{Clark}.  In this theory the uncorrelated shell model or Fermi-gas states 
are transformed by correlation operators to CB states without truncation of the 
Hilbert space.  The effective operators and interactions are matrix elements of the 
bare quantities in the CB states; they take into account the effects of 
short range correlations.  The correlation operators are chosen such that the 
nuclear interactions are relatively mild in the CB.  Observables are calculated 
using standard many-body perturbation theory methods in CB.  

Here we focus on weak interactions in nucleon matter. 
In variational calculations \cite{AP97} the nuclear matter wave functions are 
approximated with correlated states:
\beq
\label{cbwf}
\Psi_X = ({{\cal S}} {\prod_{i<j}} F_{ij}) \Phi_X ~, 
\eeq
where $\Phi_X$ are uncorrelated Fermi-gas (FG) states and $F_{ij}$ are pair 
correlation operators.  
The ${\cal S} \Pi $ denotes a symmetrized product necessary 
because the $F_{ij}$ and $F_{ik}$ do not commute. 
One can also relate uncorrelated shell model states to 
correlated states in a similar way.  The correlated states obtained from Eq. (\ref{cbwf}) 
are not 
orthogonal; we assume that they are orthonormalized using a combination of L{\"o}wdin 
and Schmidt transformations \cite{FP88} preserving the diagonal matrix elements 
of the Hamiltonian.  However, the orthonormalization corrections are of higher
order than those considered here.

Let $|X\rangle$ denote the orthonormal correlated states.  The effective 
interactions in CB perturbation theory are defined such that:
\ba
\langle X|H|Y \rangle &=& \langle \Phi_X | H_0 + H_I | \Phi_Y \rangle~,  \\ 
H_0 &=& \sum_i - \frac{\hbar^2}{2m} \nabla^2_i ~,  \\  
H_I &=& \sum_{i<j} v^{CB}_{ij} + \sum_{i<j<k} V^{CB}_{ijk} + ... ~. 
\ea
Here $H$ is the nuclear Hamiltonian containing realistic two- and 
possibly three-nucleon interactions.  Even when $H$ has only two-body 
interactions the CB $H_I$ can have three- and higher body terms.  
Since the correlated states are expected to be close to the eigenstates of $H$,
the non-diagonal matrix elements $\langle X \neq Y|H|Y \rangle$ are small.  This 
implies that the CB effective interactions can be used in perturbation expansions 
based on the Hartree-Fock approximation.  
However, the 1st order results  
are often not sufficiently accurate.  The product of pair correlation 
operators (Eq. (\ref{cbwf})) can not transform the uncorrelated states into the 
exact eigenstates of $H$.  CB calculations of the optical 
potential of nucleons in nuclear matter \cite{FFP83} including up to 2nd order 
terms in $H_I$, and of the 
response of nucleon matter to electromagnetic probes including correlated 
particle-hole rescattering \cite{FF89}, have been 
relatively successful.   In these works, as well as here, the three- and higher-body 
effective interactions are neglected.  

In the present work we use the static pair correlation operator:
\ba
F_{ij} &=& \sum_{p=1,6} f^p(r_{ij})O^p_{ij} ~,\\ 
O^{p=1,6}_{ij} &=& 1 ,~\tauij ,~\sigij ,~\tauij \sigij ,~S_{ij} ,~\tauij S_{ij} ~.  
\ea
In place of the $p=1,6$ superscripts we often use the letters $c,~\tau,~\sigma,~\sigma\tau,~t$ 
and $t\tau$ to denote the radial functions associated with these operators.  For example, 
\ba
f^{p=1,6}(r_{ij}) &\equiv& f^c_{ij},~f^{\tau}_{ij},~f^{\sigma}_{ij},~
f^{\sigma \tau}_{ij},~f^t_{ij},~f^{t \tau}_{ij} ~.
\ea
The $F_{ij}$ is 
obtained by minimizing the energy of symmetric nuclear matter at density $\rho = \rho_n 
+\rho_p $ using hypernetted and operator chain summation methods \cite{AP97,MPR02}.  
The results of the latest \cite{MPR02} variational calculations are briefly summarized in 
section VI for completeness.  
The Argonne v18 two-nucleon \cite{WSS95} and Urbana IX three-nucleon \cite{PPCPW97} 
interactions are used in these nuclear matter calculations, in studies of 
weak interactions of light nuclei \cite{ppcap,SW02}, and in the present work.   
However, improved models of $V_{ijk}$ are now available \cite{PPWC01}. 
The variational calculations of nucleon matter 
also include two spin-orbit terms in the $F_{ij}$ which are omitted here for simplicity.  
The variationally optimized $F_{ij}$ can depend upon the proton fraction $x_p$. 
However, this dependence seems to be relatively weak.  The effective interaction 
obtained from the $F_{ij}$ in symmetric nuclear matter gives a fair description of 
the spin susceptibility of pure neutron matter. 

Matrix elements of operators between CB states are generally calculated  using
cluster expansions \cite{PW79}.  We begin with the simplest, lowest order
two-nucleon cluster approximation  to study the general properties of the weak
one-body effective operators and   of the two-body interactions in  CB for
nucleon matter at densities $\rho =$  0.08, 0.16 and 0.24 fm$^{-3}$ and for
proton fraction  $x_p = \rho_p/\rho =$ 0.2, 0.3, 0.4 and 0.5.   In this density
range the contributions of clusters with  $\geq 3$ nucleons to the energy of
symmetric nuclear matter increases from 10 to 30\% of that of the 2-body \cite{MPR02};
thus the present results have only qualitative  significance.  We study the
density, proton fraction and momentum  dependence of the operators and the
interactions. 

Due to correlations and weak pair currents, the effective weak current 
operators have 2- and many-body terms in addition to 
the leading one-body term we consider here.
The lowest order (in cluster expansion) effective one-body F, GT and neutral current 
operators are calculated 
and their results are presented in sections II to V.  As expected the one-body CB 
matrix elements are smaller in magnitude than those in FG states. 
The dominant term responsible for the quenching arises from pion 
exchange interactions which change the isospins and spins of the nucleons. 
In the FG wave function, a nucleon in the single particle state $e^{i{\bf k}\cdot{\bf r}} 
\chi_{n\uparrow}$, for example, is a spin $\uparrow$ neutron with unit probability. 
This probability is reduced in the CB state by the spin-isospin correlation operators 
acting on the FG state. In contrast, the spin-isospin independent spatial 
correlations induced by the repulsive core in the two-nucleon interaction increase 
the magnitude of F, GT and NA matrix elements; however, they quench the neutron NV. 
The CB matrix elements of the charge current operators are found to have a 
rather small dependence on the matter density and $x_p$ within the range 
considered.  They depend primarily on the momentum transfer $q$, and only slightly
 on the initial or final nucleon momentum.  In addition to these, the neutral current 
 matrix elements also depend upon $x_p$. The proton NV matrix element is an exception;
it has large cancellations and depends on all of the relevant variables. 

The squares of the F and GT matrix elements in CB states are $\sim$ 0.8 and 0.75 
times those in FG states at small values of $q$.  Thus the present 
0th order (in CB $H_I$) 2-body cluster calculation predicts a quenching of low energy 
weak transitions in nuclei and nucleon matter by $\sim$ 20 to 25 \%.  
It is likely that higher order effects will further reduce 
the matrix elements and increase the quenching.  For example, the occupation 
probability of states with momenta $\alt k_F$ is $\sim$ 0.9 in CB states, and 
it decreases to $\sim$ 0.8 on including 2nd order $H_I$ corrections \cite{FP84}. 
In order to obtain quantitative results it will be necessary to include 
contributions of $\geq 3$-body clusters to the CB matrix elements neglected in this 
initial study.  This has been 
done for symmetric nuclear matter \cite{FFP83} with operator chain summation 
techniques; however, they are difficult to use in matter with $x_p \neq 0.5$.  Three-body 
cluster contributions in asymmetric matter can now be calculated using the 
recently developed matrix methods \cite{MPR02}.  

The results for the CB two-nucleon interaction are presented in Sect. \ref{CBI}. It 
gives a fair description of the spin-isospin susceptibilities of nucleon matter used 
to determine the effective interactions in the Landau-Migdal scheme \cite{PLSV01}. 
It also has the typical features of the effective interactions used in existing 
calculations of weak interactions in nucleon matter \cite{RPL99}.  If we assume
that the calculations with effective interactions are implicitly using CB states,
then their results should be reduced by a factor of $\sim$ 0.75 to take into
account the  quenching of the F and GT matrix elements by short range correlations. 
Attempts to  calculate the weak interaction rates in nucleon matter with the
effective operators  and CB interaction presented here are in progress. 

\section{Correlated Basis Fermi Matrix Element}  
\label{CBFME}

Let $|I\rangle$ and $|F\rangle$ denote the normalized correlated states obtained 
by operating on the FG states $|\Phi_{I}\rangle$ and $|\Phi_{F}\rangle $ 
by the correlation operator ${\cal S} \Pi F_{ij}$.  The 
CB Fermi matrix elements are given by: 
\begin{equation} 
\langle F| O_F |I \rangle  
= \frac{\langle \Phi_F|[{\cal S} \Pi F_{ij}] \ O_F\ [{\cal S} \Pi F_{ij}]| \Phi_I \rangle}
{\sqrt{\langle \Phi_F| [{\cal S} \Pi F_{ij}]^2 | \Phi_F \rangle 
 \langle \Phi_I|[{\cal S} \Pi F_{ij}]^2| \Phi_I \rangle}} ~, 
\label{cbme}
\end{equation}  
apart from the orthogonality corrections \cite{FP88} neglected here.  The corresponding uncorrelated, 
FG matrix element (FGME) is $\langle \Phi_F|O_F| \Phi_I \rangle $.  
It is non-zero only when the occupation numbers of the states $\Phi_I$ and $\Phi_F$ 
differ by only one nucleon, since $O_F$ is a one-body operator.  In contrast the 
CB matrix element (CBME) can be non-zero even when the occupation 
numbers of $\Phi_I$ and $\Phi_F$ differ by more than one nucleon.  However, here 
we consider only the dominant ``one-body'' CBME in which they differ 
by only one nucleon.  We define the quenching factor, $\eta$, as the 
ratio of the square of these matrix element, $|$CBME$|^2/|$FGME$|^2$. 

We assume that $\PhiI$ has full neutron and proton Fermi spheres with 
momenta $k_{Fn}$ and $k_{Fp}$, and 
\beq
\PhiF = a^{\dagger}_{{\bf k}_p \chi_p} a_{{\bf k}_n \chi_n} \PhiI ~,
\label{phif}
\eeq
where $k_n \leq k_{Fn}$ and $k_p > k_{Fp}$.  
In the absence of spin-orbit correlations, 
the Fermi matrix elements are non-zero only 
when the spin state $\chi_n = \chi_p$. 
The FGME=1 when ${\bf k}_p-{\bf k}_n = {\bf q}$.  
These conditions are also necessary for the CBME to be nonzero; however, its value  
can depend upon the matter density, proton fraction and the 
magnitudes $k_n$, $k_p$ and $q$.   

The cluster expansion of the CBME is obtained by replacing the 
correlation operators $F_{ij}$ by $1+(F_{ij}-1)$ \cite{PW79} and expanding the 
numerator and the denominator in powers of ($F_{ij}-1$).  It is convenient 
to use the $\Phi^P_I$, containing only a product of single-particle wave functions 
in which nucleons $i$ are in plane wave states with momentum ${\bf k}_i$ and 
spin-isospin $\chi_{\tau}(i)$, in place of the antisymmetric $\Phi_I$ 
and use the antisymmetric $\Phi_F$.  This is equivalent to retaining 
the antisymmetric $\Phi_I$ and $\Phi_F$ and has the advantage that we 
can associate nucleon numbers 
with the state labels in $\Phi^P_I$.  The nucleon in the state 
${\bf k}_n \chi_n$ of $\Phi^P_I$ 
is labeled ``$a$'' for active; in uncorrelated states only $a$ participates 
in the transition. All of the other nucleons in the Fermi spheres are denoted by $j$.  

\begin{figure}[htbp]
\includegraphics[height=20cm]{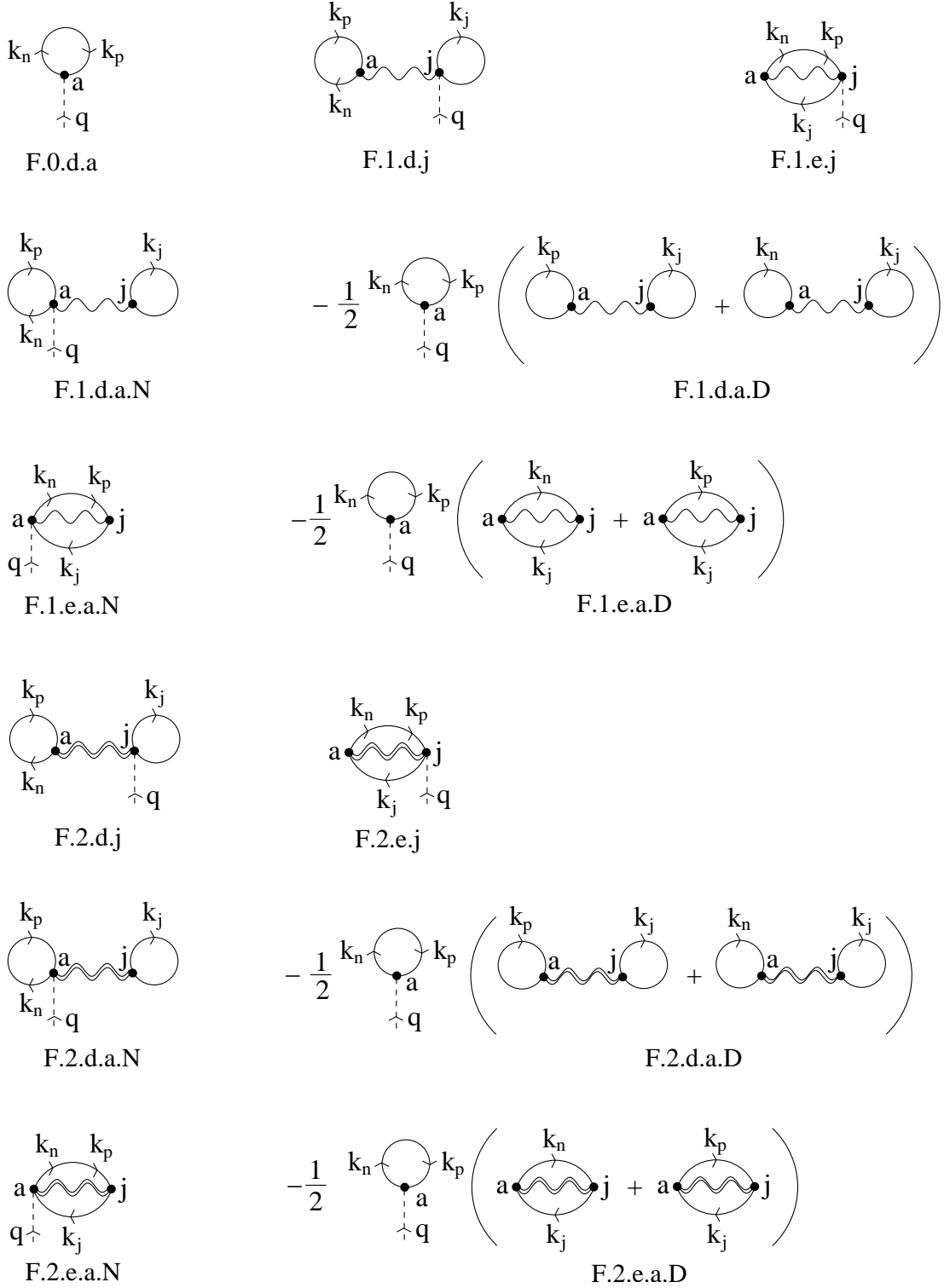}
\caption{Diagrams illustrating all of the one and two-body terms contributing to 
the Fermi CBME.}
\label{clusterd}
\end{figure}

The cluster expansion of the CBME 
is represented by diagrams as shown in Fig. \ref{clusterd}.  The terms 
in the expansion are 
labeled with $F.n.x.y$, where F stands for Fermi, $n$ is the order of the  
($F_{ij}-1$) correlations, $x=d,e$ for direct and exchange terms, and $y=a,j$ denoting 
the nucleon on which the weak interaction operates.  The dots 
in these diagrams denote nucleons, a thin line specifying
the states occupied by the nucleon in $\Phi^P_I$ and $\Phi_F$ passes through 
each dot. 
The nucleons $a$ and $j$ occupy states ${\bf k}_n$ and ${\bf k}_j$ 
in the $\Phi^P_I$, therefore lines labeled ${\bf k}_n$ and ${\bf k}_j$ originate 
from them in all diagrams.  Their termination depends upon the 
exchange pattern, since $\Phi_F$ is antisymmetric. In 
direct terms the state line ${\bf k}_j$ emerges and ends in the dot $j$ because 
the state of nucleon $j$ is unchanged.  The line
with the two labels ${\bf k}_n$ and ${\bf k}_p$ denotes the weak transition. In direct
diagrams it begins and ends in the dot $a$. In diagrams in which $a$ and $j$ are 
exchanged, the transition line 
begins at $a$ and ends in $j$, while the state line ${\bf k}_j$ begins from $j$ and ends in 
$a$.  The state and transition lines must form closed loops in all diagrams. 
The dashed line attached to nucleon $i=a$, or $j$  shows the Fermi 
operator $O_F(i)=\boldtau_i^+ e^{i {\bf q}\cdot {\bf r}_i}$.  The
$(F_{ij}-1)$ correlations are indicated by wavy lines.
We sum over the spin-isospin states $\chits (j)$ of the nucleon $j$, while those 
of $a$, $\chinu$ and $\chipu$, are specified by $\Phi_F$ (Eq. (\ref{phif})). 

The equations for $F.n.x.y$ are given below in the two-body cluster approximation 
in which $n \leq 2$. They show that 
the $F.n.x.y$ are independent of $q,k_n$ and $k_p$ when $x,y=d,a$; they 
depend only on $q$ when $x,y=d,j$; and only on $k_n$ and $k_p$ in exchange 
diagrams ($x=e$).  We also give a simple explanation of the important 
$F.2.d.a$ term responsible for much of the quenching.  
The standard 2nd order perturbation theory calculation of the direct 
contributions to the Fermi matrix element is reviewed in Appendix A.  One can easily 
identify the analogues of $F.n.d.y$ in that familiar theory and obtain relations 
between the present approach and that of Arima and coworkers \cite{Arima87}. 
The perturbation theory assumes that the forces are weak, but in reality 
we cannot expand in powers of the strong, bare 2-nucleon interaction $v_{ij}$. 
However, we hope that standard perturbation theory can be used in CB 
with the effective operators and interactions described here, as mentioned in the 
introduction. 

The leading 0th order term is given by: 
\beq 
F.0.d.a = FGME = \int d^3r~e^{i({\bf k}_n+{\bf q}-{\bf k}_p)\cdot {\bf r}}  
\langle \chipu (a) | \boldtau^+(a) | \chinu (a) \rangle = 1 ~.
\label{FGME}
\eeq 
The momentum conserving delta function
$\delta^3({\bf k}_p-{\bf k}_n-{\bf q})$ and the $\chinu = \chipu$ spin constraint are implied 
here as well as in all terms of the
expansion given below.  There are no other $0$th order terms. 

The 1st order direct term with $O_F(j \neq a)$ is given by:
\ba
F.1.d.j &=& \sum_{j} \int d^3r_{aj}~e^{-i {\bf q}\cdot{\bf r}_{aj}} \langle \chipu (a) \chits (j) | 
\{ \boldtau^+_j~,(F_{aj}-1) \}|\chinu (a) \chits (j) \rangle \nonumber \\
&=& \rho \int d^3r \ e^{-i {\bf q}\cdot{\bf r}} \ 2 \ f^{\tau} (r)~.   
\label{f1dj}
\ea
All spin dependent terms in $F_{aj}$ give zero contribution on summing over the spin 
states of nucleons $j$, and the factor of 2 in the above equation comes from: 
\beq
\{\boldtau^+_j~,~\boldtau_j \cdot \boldtau_a \} = 2 \boldtau^+_a ~.
\eeq
From now on the $aj$ subscripts on ${\bf r}$ and $F$ will be dropped for brevity, 
and the $r$ dependence of the $f^p$'s will be implicit.

The contribution of $F.1.e.j$ is given by: 
\ba
F.1.e.j &=&\sum_{j} \int d^3r~e^{i ({\bf k}_n-{\bf k}_j)\cdot{\bf r}} 
\langle \chipu (a) \chits (j) |e_{aj}  
\{ \boldtau^+_j~,(F-1) \}|\chinu (a) \chits (j) \rangle \nonumber \\
&=& - \int d^3r~e^{i {\bf k}_n\cdot{\bf r}} \left[ \rho_n~\ell_n(r) (f^c-1+3f^{\sigma}) 
+\rho_p~\ell_p(r) (f^{\tau}+3f^{\tau \sigma}) \right] , 
\label{f1ej}
\ea
where $e_{ij}$ is the spin-isospin exchange operator: 
\beq
e_{ij} = - \frac{1}{4}(1+\tauij)(1+\sigij) ~, 
\eeq
and the Slater functions $(N=n,p)$ are:
\beq
\ell_N(r) = \frac{2}{\rho_N} \int \frac{d^3k}{(2\pi)^3} \theta(k_{FN}-k)~e^{i {\bf k}\cdot {\bf r}} 
= 3[sin(k_{FN}r)- k_{FN}r \ cos(k_{FN}r)]/(k_{FN}r)^3 ~. 
\eeq
The algebra of the operators $O^{p=1,6}_{aj}$ described in Ref. \cite{PW79} is very useful 
in evaluating these contributions. 

The 2-body terms with $O_F(a)$ have contributions from the numerator of the
matrix element, Eq. (\ref{cbme}), as well as normalization corrections
introduced through the expansion of the denominator.   We denote these by
$F.1.x.a.N$ and $F.1.x.a.D$ respectively.   In Fig. \ref{clusterd} the
denominator contributions are shown as products of two diagrams.  
The 1st order direct terms with $O_F(a)$ cancel:
\beq
F.1.d.a = F.1.d.a.N + F.1.d.a.D = 0 ~, 
\label{f1da}
\eeq
while for the exchange terms we obtain: 
\ba
F.1.e.a &=& F.1.e.a.N + F.1.e.a.D  \\ 
F.1.e.a.N &=& \sum_j \int d^3r~e^{-i ({\bf k}_j-{\bf k}_p ) \cdot{\bf r}}
 \langle \chipu (a) \chits (j) | 
e_{aj} \{ \boldtau^+_a~,(F-1) \}|\chinu (a) \chits (j) \rangle \nonumber \\
&=& - \int d^3r~e^{i {\bf k}_p \cdot{\bf r}}
[ \rho_p \, \ell_p(\gfa -1 + 3 \gfc) + \rho_n \, \ell_n(\gfb + 3 \gfd)]~, 
\label{f1ean}
\ea
\ba
F.1.e.a.D &=& \sum_j -\int d^3r \Big( e^{-i ({\bf k}_j-{\bf k}_n ) \cdot{\bf r}}
 \langle \chinu (a) \chits (j) | 
e_{aj} (F-1)|\chinu (a) \chits (j) \rangle \nonumber \\ 
&& + \  e^{-i ({\bf k}_j-{\bf k}_p ) \cdot{\bf r}} \langle \chipu (a) \chits (j) | 
e_{aj} (F-1)|\chipu (a) \chits (j) \rangle \Big) \nonumber \\
&=& \frac{1}{2} \int d^3r \Big([\gfa -1 + 3 \gfc + \gfb + 3 \gfd]
[e^{i {\bf k}_n \cdot{\bf r}} \rho_n \, \ell_n + e^{i {\bf k}_p \cdot{\bf r}}
\rho_p \, \ell_p]   \nonumber \\
&&+ 2 [\gfb + 3 \gfd][e^{i {\bf k}_n \cdot{\bf r}} \rho_p \, \ell_p
+ e^{i {\bf k}_p \cdot{\bf r}}\rho_n \, \ell_n]\Big)~.
\label{f1ead}
\ea

For calculating the 2nd order terms, it is convenient to define: 
\ba 
F &=& 1 + F^0 + F^1 \boldtau_a \cdot \boldtau_j ~, \\ 
F^0 &=& f^c - 1 + f^{\sigma}\boldsigma_a \cdot 
\boldsigma_j  + f^t S_{aj}~,  \\  
F^1 &=& f^{\tau} + f^{\sigma \tau}\boldsigma_a \cdot 
\boldsigma_j  + f^{t \tau} S_{aj}~.
\ea 
Only the spin independent parts of the
products of the above $F^0$ and $F^1$ contribute  to the second order 
diagrams.  These are called the $C-$ parts in Ref. \cite{PW79}.  We define: 
\ba 
C^{IJ}_d &=& C[F^IF^J]~, \\  
C^{IJ}_e &=& C[(1+ \boldsigma_a \cdot \boldsigma_j ) F^IF^J] ~.
\ea
The expressions for $C^{IJ}_d$ and $C^{IJ}_e$ in terms of the correlation
functions, $f^p$, are given in Appendix \ref{appen}.

There is no contribution from the denominator to the terms $F.2.x.j$. 
These are given by:
\ba
F.2.d.j &=& \sum_j \int d^3r ~ e^{-i {\bf q} \cdot {\bf r}}
 \langle \chipu (a) \chits (j) | (F-1) \boldtau^+_j (F-1)|
 \chinu (a) \chits (j) \rangle  \nonumber \\
 &=& \rho \int d^3r e^{-i {\bf q} \cdot {\bf r}} ~2 \big[ C^{11}_d  + C^{01}_d \big] ~,
 \label{f2dj}
\ea
\ba 
F.2.e.j &=& \sum_j \int d^3r ~ e^{i ({\bf k}_n-{\bf k}_j) \cdot {\bf r}}
\langle \chipu (a) \chits (j) | e_{aj} (F-1) \boldtau^+_j (F-1)|
 \chinu (a) \chits (j) \rangle  \nonumber \\
 &=& -\frac{1}{2} \int d^3r e^{i {\bf k}_n \cdot {\bf r}} \big[ \rho_n \, \ell_n
 (C^{00}_e - C^{11}_e ) + 2~\rho_p \, \ell_p
 ( C^{11}_e + C^{01}_e)\big] ~. 
\label{f2ej}
\ea
The sum: 
\ba
F.2.d.a &=& F.2.d.a.N + F.2.d.a.D ~=  \nonumber \\ 
&\sum\limits_j& \int d^3r \langle \chipu (a) \chits (j) | (F-1)\boldtau^+_a (F-1) - \frac{1}{2} 
\{ \boldtau^+_a,~(F-1)^2 \}| \chinu (a) \chits (j) \rangle  \nonumber \\
&=& \rho \int d^3r (-4 C^{11}_d) ~. 
\label{f2da}
\ea
Note that only the $F^1 \boldtau_a \cdot \boldtau_j$, 
which does not commute with the $\boldtau^+_a$ operator, 
contributes to this sum.  

The results presented in the next subsection show that the above term gives the 
largest contribution to the quenching of the Fermi matrix element in matter.  This 
term simply takes into account the probability for nucleon $a$ to be a 
neutron in the initial and a proton in the final state. In the uncorrelated 
product state, $|\Phi^P_I \rangle $, nucleon $a$ is $n\uparrow$; but in 
the correlated product state, ${\cal S}\Pi F_{ij} | \Phi^P_I \rangle $, it can be 
in other nucleon states.  We refer to nucleon $a$ in the correlated state as 
a ``quasi-nucleon''.  
The probability that it is a neutron is given by:
\begin{equation} 
P_I(a=n)= \frac{\langle \Phi_I|[{\cal S} \Pi F_{ij}] \ \frac{1}{2}(1-\tau^z_a) 
\ [{\cal S} \Pi F_{ij}]| \Phi^P_I \rangle}
{\langle \Phi_I|[{\cal S} \Pi F_{ij}]^2| \Phi^P_I \rangle}  
\label{pian}
\end{equation}  
We use the cluster expansion to calculate this probability.  The 0th order, one-body 
term is unit, and the two-body 2nd order direct term is:
\ba
&-& \frac{1}{2} 
\sum_j \int d^3r \langle \chinu (a) \chits (j) | (F-1)\tau^z_a (F-1) - \frac{1}{2}
\{ \tau^z_a,~(F-1)^2 \}| \chinu (a) \chits (j) \rangle  \nonumber \\
&=& \rho_p \int d^3r (-4 C^{11}_d) ~.
\label{pian2}
\ea
The two-body 1st order direct terms cancel as in Eq. (\ref{f1da}). Neglecting the exchange terms 
we obtain the direct part:
\beq
P_I(a=n,~d) = 1 + \rho_p \int d^3r (-4 C^{11}_d) ~.
\label{pian3}
\eeq
In a similar way, the direct part of the 
probability for the active quasi-nucleon, $a$, to be a proton in the final state is given by: 
\beq
P_F(a=p,~d) = 1 + \rho_n \int d^3r (-4 C^{11}_d) ~.
\label{pfap}
\eeq
Hence
\beq
1+F.2.d.a = P_I(a=n,~d)~P_F(a=p,~d) ~ , 
\eeq
neglecting the terms of order $(C^{11}_d)^2$.  

The probabilities for the active quasi-nucleon to be in the initial 
spin isospin states
$\uparrow,~\downarrow$ $n, p$ have been calculated keeping only the direct terms, 
at the three densities 
for $x_p=0.5$.  These are given in Table \ref{table2}.
In one-body Fermi transitions these are also the probabilities for the active 
quasi-nucleon to be a spin $\uparrow,~\downarrow$ $p,n$ in the final state.

The 2nd order exchange term:
\beq
F.2.e.a = F.2.e.a.N + F.2.e.a.D ~, 
\eeq
has contributions from both $F^1$ and $F^0$. They are given by:
\ba
F.2.e.a.N &=& \sum_j \int d^3r ~ e^{i ({\bf k}_p - {\bf k}_j) \cdot {\bf r}}
\langle \chipu (a) \chits (j) | e_{aj} (F-1) \boldtau^+_a (F-1)|
 \chinu (a) \chits (j) \rangle  \nonumber \\
&=& -\frac{1}{2} \int d^3r e^{i {\bf k}_p \cdot {\bf r}} \big[ \rho_p \, \ell_p
(C^{00}_e-C^{11}_e) 
+ 2 \rho_n \,\ell_n
( C^{11}_e + C^{10}_e)\big]~,
\label{f2ean}
\ea
\ba
F.2.e.a.D &=& \sum_j -\frac{1}{2}\int d^3r \Big( e^{-i ({\bf k}_j-{\bf k}_n ) \cdot{\bf r}}
 \langle \chinu (a) \chits (j) | 
e_{aj} (F-1)^2|\chinu (a) \chits (j) \rangle \nonumber \\ 
&& + \  e^{-i ({\bf k}_j-{\bf k}_p ) \cdot{\bf r}} \langle \chipu (a) \chits (j) | 
e_{aj} (F-1)^2|\chipu (a) \chits (j) \rangle \Big) \nonumber \\
&=& \frac{1}{4} \int d^3r \Big(  (-4 C^{11}_e + 4 C^{10}_e 
)(e^{i {\bf k}_n \cdot{\bf r}}\rho_p \,\ell_p  
 + e^{i {\bf k}_p \cdot{\bf r}}\rho_n \, \ell_n) \nonumber \\
&& + (C^{00}_e + C^{11}_e + 2 C^{10}_e ) 
(e^{i {\bf k}_n \cdot{\bf r}}\rho_n \, \ell_n  
 + e^{i {\bf k}_p \cdot{\bf r}}\rho_p \, \ell_p)
\Big)~.
\label{f2ead}
\ea

\subsection{Results for Fermi Matrix Element}
\label{rfme}
\begin{figure}[htbp]
\includegraphics[height=15cm, angle=-90]{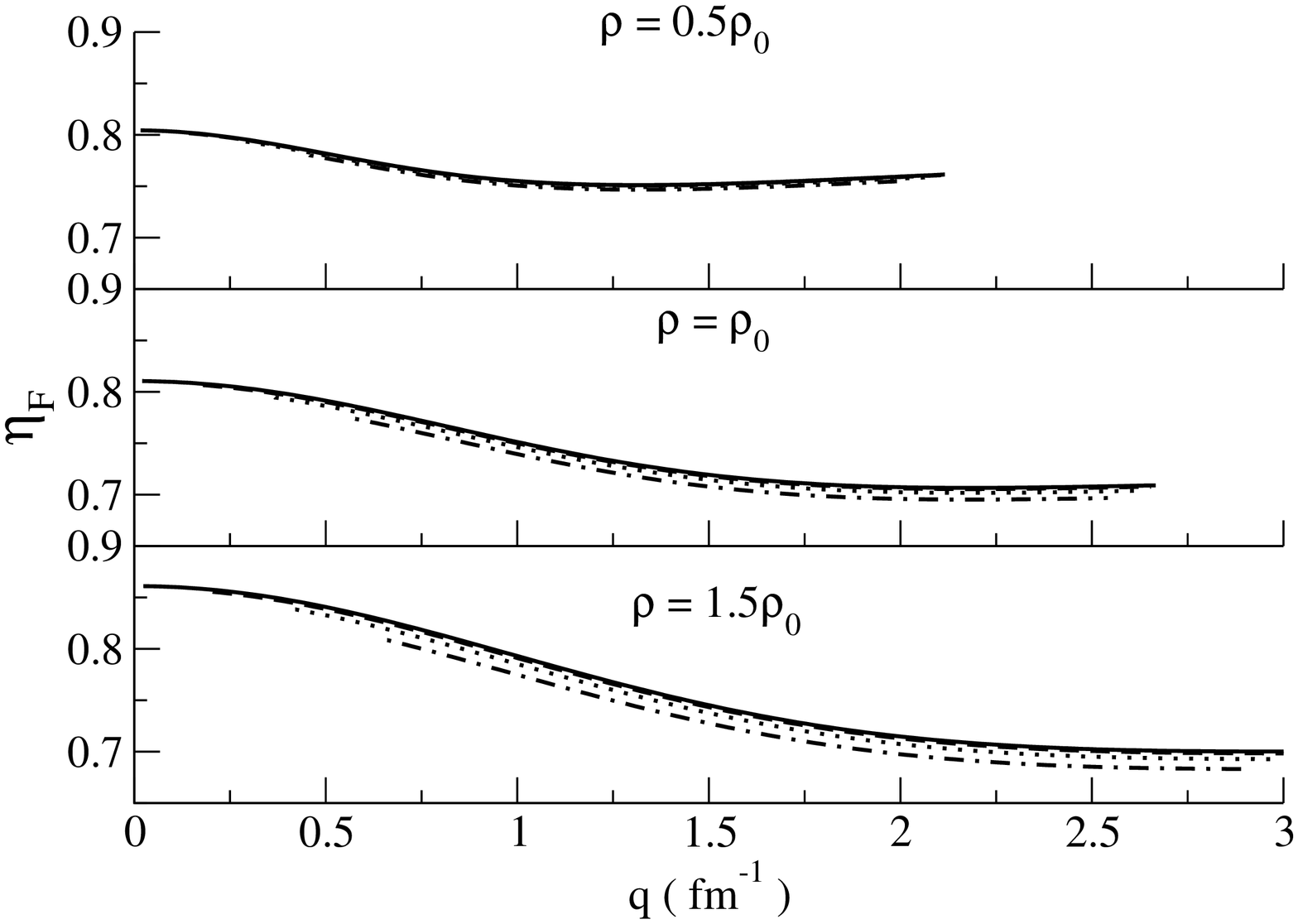}
\caption{$\eta_F$ as a function of $q$ and proton fraction
$x_p$ for $k_N=k_{FN}$.  The solid, dashed, dotted and dash-dot lines 
show results for $x_p=0.5,~0.4,~0.3,$ and 0.2.} 
\label{fallrho}
\end{figure}

The Fermi matrix elements have been calculated using correlation functions
obtained in Ref. \cite{AP97} by minimizing the energy of symmetric nuclear matter using the 
Argonne-v18 and Urbana IX 2- and 3-nucleon interactions.  
In Fig. \ref{fallrho} we present the results for $\eta_F$, the square of the 
Fermi CBME (Eq. (\ref{cbme})), 
for $k_n = k_{Fn}$ and $k_p = k_{Fp}$.  

When $x_p < 0.5$ the  total isospin $T_I$ of the state $|I\rangle$ is $(N-Z)/2$, while
that of  $|F\rangle$ is $(N-Z)/2-1$.   In the case of symmetric nuclear matter the
$T_I=0$, while $T_F=1$.  Thus the  calculated matrix elements are between states with
$\Delta T = 1$.   The Fermi matrix elements for $q=0$, between isobaric analogue states
having the  same $T$ and $T_{zF} = T_{zI}\pm 1$ are given by $(T\mp T_{zI})(T \pm
T_{zI}+1)$ in both  correlated and uncorrelated states.  We will not discuss 
$\Delta T = 0$ Fermi ME in this paper.

The variation of $\eta_F$ with proton fraction is less than 
$3\%$ at all densities calculated.  However, the proton fraction limits the allowed values
of $q$ through the momentum conservation relation: ${\bf q} = {\bf k}_p -
{\bf k}_n$.  The variation with total density is also small within the considered range.  
This suggests that we can approximate $|$CBME$|^2$ by a function of 
$\rho$ and $q$.  
In the small $q$ region, $q \lesssim 0.5$ fm$^{-1}$, 
it can be well represented by the quadratic: 
\beq
\eta_F = \eta_F(q \rightarrow 0) + \alpha_F(\rho) \, q^2 ~. 
\label{quad}
\eeq
We have fit the calculated values for symmetric nuclear matter and
the results are given in Table \ref{table1}.

\begin{figure}[htbp]
\includegraphics[height=15cm, angle=-90]{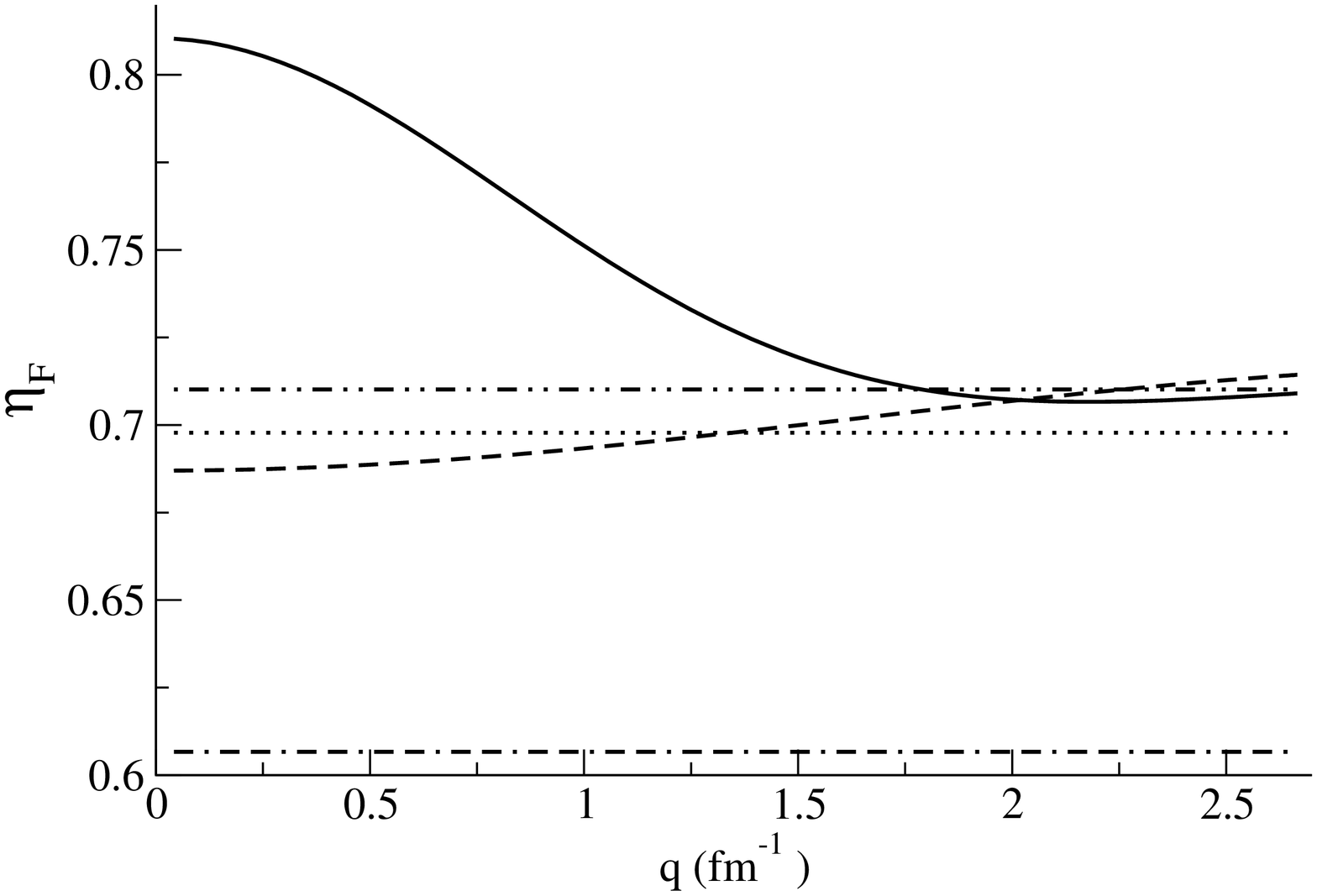}
\caption{Contributions to $\eta_F$ for $k_N=k_{FN}$ and
$\rho = \rho_0$.  The dash-double dot line includes all of the $q$-independent terms, while 
the solid line shows the full result.  See text for description of other 
curves.}
\label{fparts}
\end{figure}

Fig. \ref{fparts} shows the contributions of each
term in the cluster expansion of the Fermi matrix element in matter 
at density $\rho_0$ and $x_p=0.5$.  
The $F.n.x.a$ and $F.n.e.j$ terms give contributions 
that are independent of ${\bf q}$ as can be seen from Eqs. (\ref{f1ej}), 
(\ref{f1ean}), (\ref{f1ead}), (\ref{f2ej}), (\ref{f2da}), (\ref{f2ean}) and
(\ref{f2ead}).  
The dominant contribution to the quenching of the Fermi CBME comes from 
$F.2.d.a$; $|1+ F.2.d.a|^2=0.7$ is shown by the dotted line in 
Fig. \ref{fparts}.  As discussed in the previous subsection this result 
can be interpreted in terms of the probabilities for the active quasi-nucleon $a$ 
to be a neutron in the initial and a proton in the final CB states. 

The exchange terms, $F.n.e.a$, contribute an additional 
$\sim$ 0.1 to the $q$-independent quenching; 
$| \sum_{n,x} F.n.x.a |^2=0.61$ is shown by the 
double dash-dot line.  This additional quenching is mostly canceled by 
the $F.n.e.j$ terms, as shown by the dash-double dot line;
$|\sum_{n,x}F.n.x.a + F.n.e.j|^2=0.71$.  

The $F.n.d.j$ terms, given by Eqs. (\ref{f1dj}) and (\ref{f2dj}), 
introduce the $q$-dependence.  Of these, the 2nd order $F.2.d.j$ is dominant 
as can be seen from the dashed line, which includes only $F.1.d.j$ and 
all of the $q$-independent terms. The full line gives the 
square of the total matrix element including $F.2.d.j$.  

\begin{figure}[htbp]
\includegraphics[height=15cm, angle=-90]{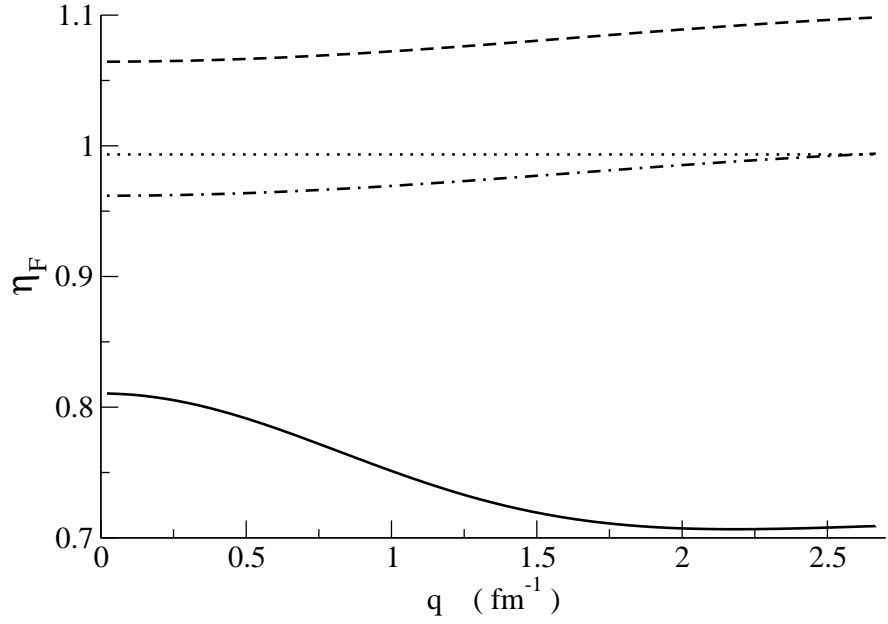}
\caption{Correlation dependence of $\eta_F$ for $k_N=k_{FN}$ and
 $\rho = \rho_0$.   The dashed line 
shows results with $\gfd=\gff=0$, and in addition, $f^c=1$ for the dash-dot 
line. The dotted line shows $|\sum_{n,x} F.n.x.a|^2$ when $\gfd=\gff=0$. The solid
line gives the full result.} 
\label{fcorparts}
\end{figure}

The contributions of the various correlations 
to the CBME are shown in Fig. \ref{fcorparts}.  The 1st and 2nd order terms are dominated by the 
$\gfd (r_{ij})\sigij \tauij$ and $\gff (r_{ij})S_{ij} \tauij$ correlations 
induced mainly be the OPEP.  After setting 
$\gfd = \gff =0$ the $|\sum_{n,x} F.n.x.a|^2$ becomes essentially 1 as 
shown by the dotted line in Fig. \ref{fcorparts}.  The full CBME exceeds unity 
in this case (see the dashed line) via the contributions of $f^c-1$ 
correlations to $F.n.x.j$.  The dash-dot line shows 
$\eta_F$ obtained by further setting $\gfa = 1$.  It is fairly close to 
one showing that the $\gfb,\gfc$ and $\gfe$ correlations have small 
effects.

\begin{figure}[htbp]
\includegraphics[height=15cm, angle=-90]{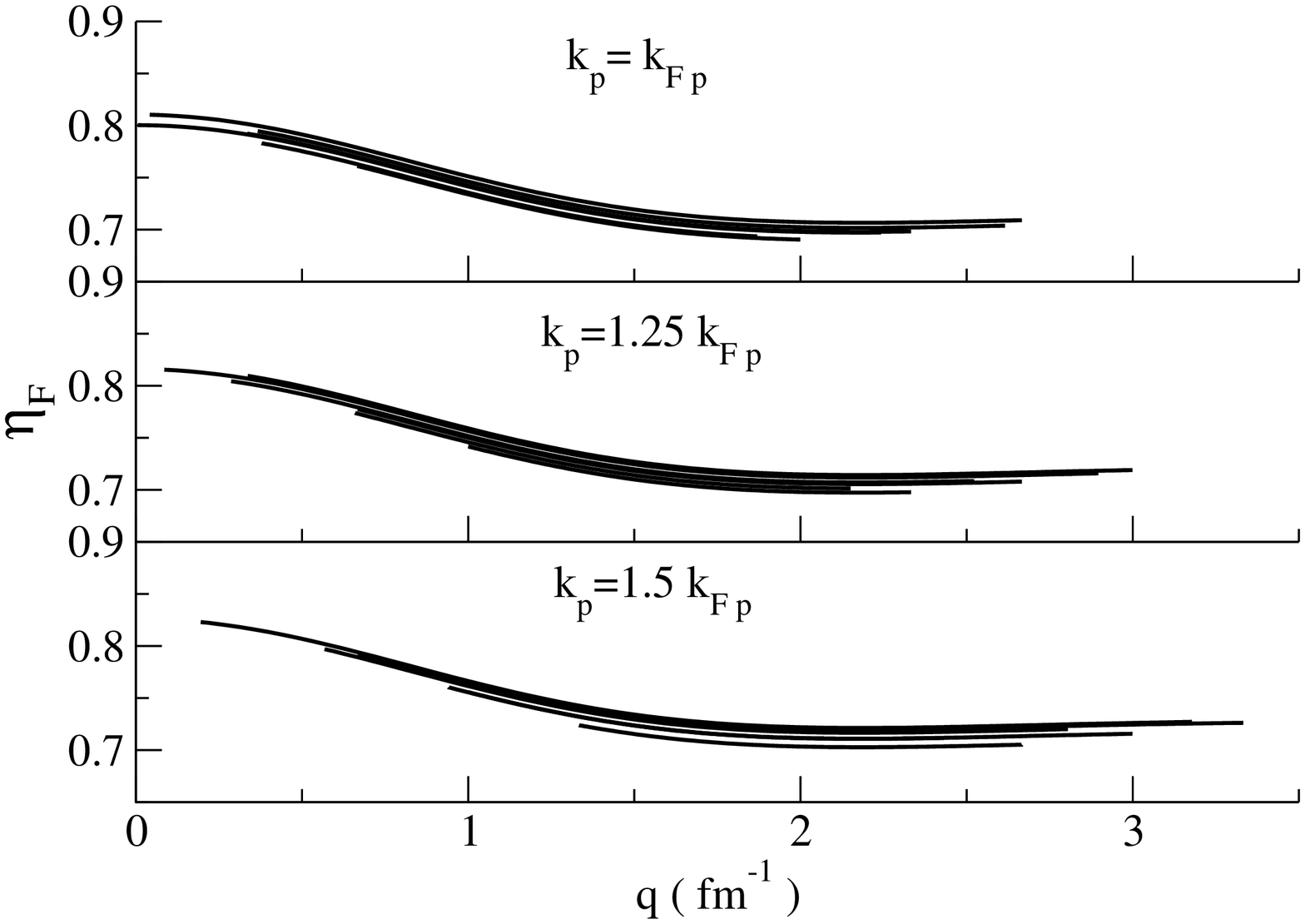}
\caption{Dependence of $\eta_F$ on the initial ($k_n$) 
and final ($k_p$) momenta for $\rho = \rho_0$.  
Each set contains six lines depicting the results for $k_n=(.5,~.75,~1)k_{Fn}$, 
and $x_p=0.3$ and 0.5 for the indicated value of $k_p$.} 
\label{knkpdep}
\end{figure}

The Fermi CBME, calculated in the two-body cluster approximation does not 
depend significantly on the magnitudes of initial and final nucleon 
momenta.  The dependence on $k_{Fn}-k_n$ and $k_p-k_{Fp}$ is illustrated in 
Fig. \ref{knkpdep}.  It shows $\eta_F$ for $\rho = \rho_0$, 
$x_p = 0.5$ and 0.3, $k_n = 1,~0.75,~0.5~k_{Fn}$ and 
$k_p = 1,~1.25,~1.5~k_{Fp}$ as a function of $q$.  The results for 
the 18 possible combinations of $x_p$, $k_n$ and $k_p$ values 
differ by less than 0.03. 

\section{Correlated Basis Gamow-Teller Matrix Element}  
\label{GTME}

The procedure for the calculation of the GT matrix element is similar
to that discussed previously in Section \ref{CBFME}.  We therefore discuss only
the differences and give the final expressions.  The
operator,
$\mathbf{O}_{GT}$, is an axial vector and it is convenient to express
its matrix element using the following two axial vectors:
\beq
\siga = \langle \chi_p(a) | \, \boldsigma(a) \ \boldtau^{+}(a)\,  | \chi_n(a) \rangle
\eeq
and
\beq
\At = 3 \ \hat{\bf r}_{aj} \ \siga \cdot \hat{\bf r}_{aj} - \siga ~ ,
\eeq
obtained from the tensor correlations between nucleons $a$ and $j$. 
Note that $\At$ depends upon $\hat{\bf r}_{aj}$. 

We assume that $\chi_n$ in Eq. (\ref{phif}) is spin up 
and sum the square of the GT matrix element for 
the two final states with $\chi_p = \uparrow , \downarrow$ denoted by $|F\uparrow \rangle$ 
and $|F\downarrow \rangle$. 
In FG we get contributions only via 
the operator $\widetilde{\boldsigma_a} = \boldsigma_a \boldtau_a^+$; only $\sigma_{z}(a)$ 
contributes to the FGME with $\chi_p = \uparrow $, while 
$\sigma_{x}(a)$ and $\sigma_{y}(a)$ give the 
GT FGME for $ \chi_p = \downarrow $.  However, in CB the 
$\widetilde{{\bf A}_t}$ induces transitions that are forbidden in FG states.  

The terms in the cluster expansion of the GT CBME are denoted by 
$GT.n.x.y$ as in the last section.  The ratio $g_A$ of the axial to vector 
coupling constants is omitted from the $GT.n.x.y$ for brevity.  We obtain: 
\ba 
GT.0.d.a & = & \siga
\ea
\ba
GT.1.d.j &=& \rho \int \! \! d^3r \, e^{-i \mathbf{q}\cdot\mathbf{r}} 
\ 2\, \big(\gfd \, \siga + \gff \, \At\big) \\
GT.1.e.j &=& -\int \! \! d^3r \, e^{i \mathbf{k}_n\cdot\mathbf{r}} \Big\{ 
\rho_p \, \ell_p (\gfb + 3 \gfd) \, \siga \nonumber \\
&&+ \rho_n \, \ell_n \Big[(\gfa - 1 + \gfc + 2 \gfd) \, \siga + 
(\gfe - \gff) \,\At \Big] \Big\}
\ea
\ba
GT.1.e.a.N &=&-\int \! \! d^3r \, e^{i \mathbf{k}_p\cdot\mathbf{r}} \Big\{ \rho_n \,
 \ell_n (\gfb + 3 \gfd) \, \siga \nonumber \\
&&+ \rho_p \, \ell_p \Big[(\gfa - 1 + \gfc + 2 \gfd) \,\siga + (\gfe - \gff) \,
\At \Big] \Big\} \\
GT.1.e.a.D &=&  \siga F.1.e.a.D
\ea
\ba
GT.2.d.j&=& 2 \rho \int \! \! d^3r e^{-i \mathbf{q}\cdot\mathbf{r}}\Big[
\big(\gfTTjdc+\gfoTjdc \big) \,\siga +\big(\gfTTjdt +\gfoTjdt \big) \, \At
\Big]\\
GT.2.e.j &=& -\frac{1}{2} \int \! \! d^3 r \, e^{i \mathbf{k}_n\cdot\mathbf{r}}
\Big\{ \rho_n \, \ell_n\Big[\big(\gfoojec - \gfTTjec -\gfTojec +\gfoTjec
\big) \, \siga \nonumber \\
&&+ \big(\gfoojet - \gfTTjet -\gfTojet +\gfoTjet\big) \, \At\Big] 
+ 2 \rho_p \, \ell_p \Big[\big(\gfTTjec + \gfoTjec\big) \, \siga \nonumber \\
&&+\big(\gfTTjet  + \gfoTjet \big) \, \At\Big]\Big\}
\ea
\ba
GT.2.d.a &=& \rho \int \! \! d^3 r 
\big(\gfooadc - \gfTTadc -C^{00}_d - 3 C^{11}_d) \,\siga \\
GT.2.e.a.N&=& -\frac{1}{2} \int \! \! d^3 r \, e^{i \mathbf{k}_p\cdot\mathbf{r}}
\Big\{ \rho_p \, \ell_p\Big[\big(\gfooaec - \gfTTaec +\gfToaec -\gfoTaec
\big) \, \siga \nonumber \\
&&+ \big(\gfooaet - \gfTTaet +\gfToaet -\gfoTaet\big) \, \At\Big] 
+ 2 \rho_n \, \ell_n \Big[\big(\gfTTaec + \gfToaec 
\big) \, \siga \nonumber \\
&&+\big(\gfTTaet + \gfToaet \big) \, \At\Big]\Big\} \\
GT.2.e.a.D &=& \siga F.2.e.a.D
\ea
The coefficients $F^{IJ, \sigma}_{d,y}$ ($y=a,j$) and $F^{IJ, A}_{d,y}$ are defined as the
$\boldsigma_a$ and ${\bf A}_t$ parts of the operator
$F^I \boldsigma_y F^J$:
\beq
F^I \boldsigma_y F^J = F^{IJ, \sigma}_{d,y} \boldsigma_a + F^{IJ, A}_{d,y}{\bf A}_t
+ \mathrm{terms \ linear \ in \ } \boldsigma_j ~.
\eeq
The remaining parts linear in $\boldsigma_j$ do not contribute after summing over $\boldsigma_j$.
The $F^{IJ, \sigma}_{e,y}$ and $F^{IJ, A}_{e,y}$ are the corresponding 
parts of the operator $(1+\boldsigma_a \cdot \boldsigma_j)F^I \boldsigma_y F^J$, and the 
expressions for $F^{IJ, \sigma}_{x,y}$ and 
$F^{IJ, A}_{x,y}$ are given in Appendix \ref{appen}.

As in the Fermi case, the 2nd order direct diagrams, $GT.2.d.a$ can be
interpreted in terms of quasi-nucleon probabilities.  The $GT.2.d.a$ has contributions from 
the $\siga$ only.  When the final proton has spin $\uparrow$ only the 
$\sigma_a^z$ term contributes.  We consider this simple 
case for illustration. In this case, $(1+GT.2.d.a)$ represents the 
probability that the active quasi-nucleon has $\sigma_{a}^z \, \tau_{a}^z
= -1$ in the initial state and $+1$ in the final state.
In FG states these are unit probabilities. 
We use the cluster expansion to calculate them in CB states. The 0th order
terms $= 1$, and the two-body 2nd order direct terms are given by:
\ba
&\mp& \frac{1}{2}  \sum_j \int d^3r \langle \chinu (a) \chits (j) | (F-1)\sigma_a^z
\,\tau^z_a (F-1) - \frac{1}{2} \{\sigma_a^z \, \tau^z_a,~(F-1)^2 \}| \chinu (a)
\chits (j) \rangle  \nonumber \\  
&=& \rho \, \frac{1}{2} \int d^3r \,(\gfooadc -
\gfTTadc - C^{00}_d - 3 C^{11}_d) \nonumber \\
&&\mp \, (\rho_p-\rho_n)\frac{1}{2} \int d^3r \, (
\gfToadc + \gfoTadc + 2 \gfTTadc - 2 C^{10}_d + 2 C^{11}_d) ~,
\label{pist}
\ea
where the upper and lower signs correspond to the initial and final states
respectively.   The 1st order direct terms cancel as in Eq. (\ref{f1da}). 
Neglecting the exchange terms and those of order $C^{IJ}_d C^{MN}_d$ and
$F^{IJ,\sigma}_{d,a} F^{MN,\sigma}_{d,a}$ we obtain:
\ba
P_I(\sigma_{a}^z \, \tau_{a}^z = -1,~d) ~
P_F(\sigma_{a}^z \, \tau_{a}^z = 1,~d) &=& 1+
\rho \, \int d^3r \,(\gfooadc - \gfTTadc - C^{00}_d - 3 C^{11}_d) \nonumber
\\
&=& 1+[GT.2.d.a]_z~.
\ea
The $\sigma_{a}^z \, \tau_{a}^z = -1$ probability is the sum of the $n\uparrow$ 
and $p\downarrow$ probabilities listed in Table \ref{table2}. 

\subsection{Results of Gamow-Teller Matrix Element}
\label{ssgt}

The tensor correlations lead to a dependence of the GT CBME 
on the direction of the spin quantization axis through the $\At$
terms.  
We therefore do not discuss the CBME for spin up and down final
states individually.  The sum of $|$CBME$|^2$ over the final two spin states determines the 
transition rates and is independent 
of the chosen axis.  This sum equals 3 for FGME.  In the following we report 
results for: 
\beq 
\eta_{GT}
\equiv \frac{1}{3}(|\langle F \uparrow |\mathbf{O}_{GT}| I \rangle|^2+ |\langle
F \downarrow |\mathbf{O}_{GT}| I \rangle|^2) 
\eeq

\begin{figure}[htbp]
\includegraphics[height=15cm, angle=-90]{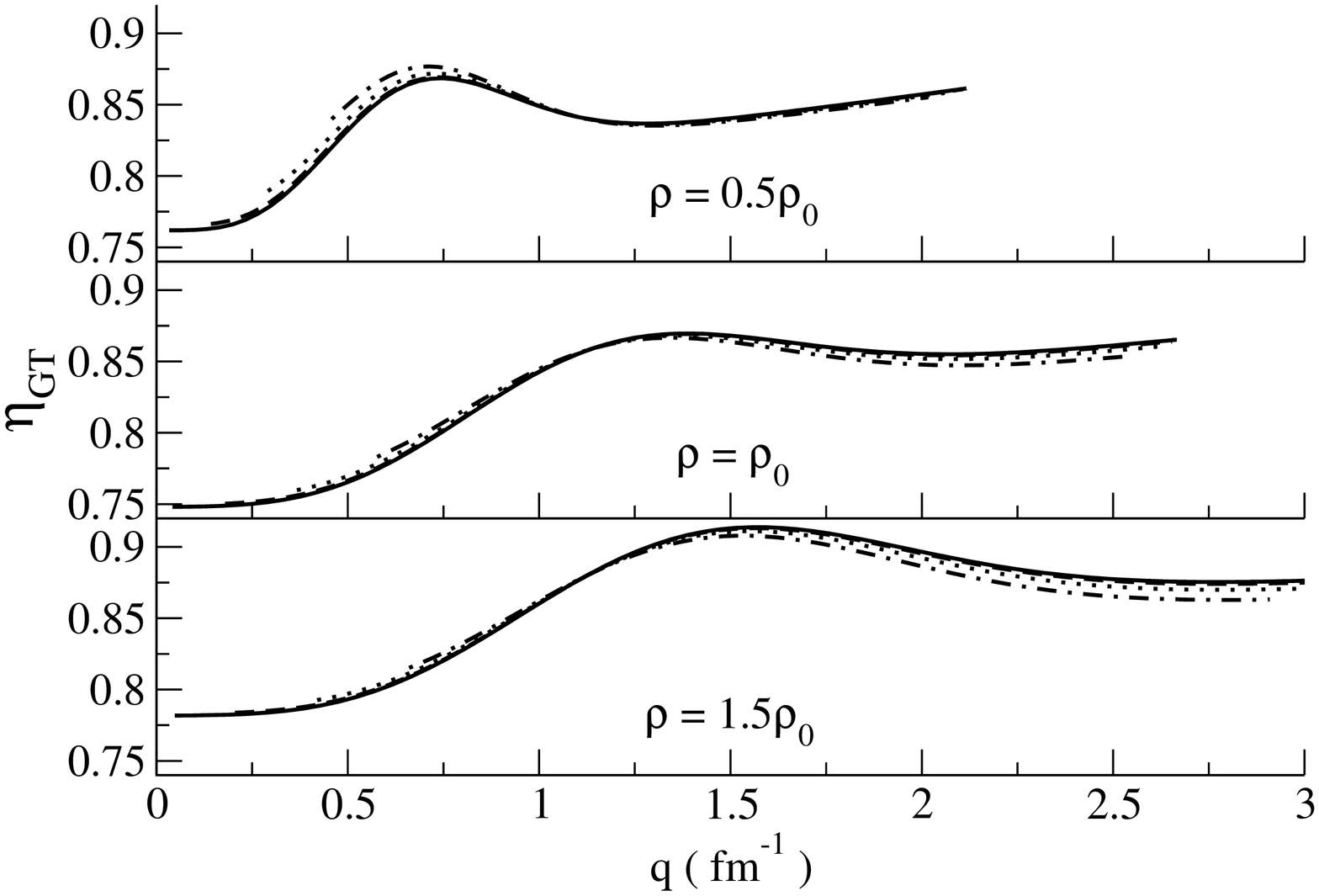}
\caption{$\gtav$ as a function of $q$ and proton fraction
$x_p$ for $k_N=k_{FN}$.  The solid, dashed, dotted and dash-dot lines 
show results for $x_p=0.5,~0.4,~0.3,$ and 0.2.} 
\label{gtallrho}
\end{figure}

The $\gtav$  has been calculated using the correlation
functions as described in Section \ref{rfme} and the results 
for $k_N=k_{FN}$ are plotted in Fig. \ref{gtallrho}.  As in 
the Fermi case, the variation of $\gtav$ due to changes in 
proton fraction is less than $3\%$, but 
it has more $q$ dependence. The 
quadratic fit (Eq. (\ref{quad})) is still valid up to $q \sim 0.5$ fm$^{-1}$, 
and its parameters are given in Table \ref{table1}.

\begin{figure}[htbp] 
\includegraphics[height=15cm, angle=-90]{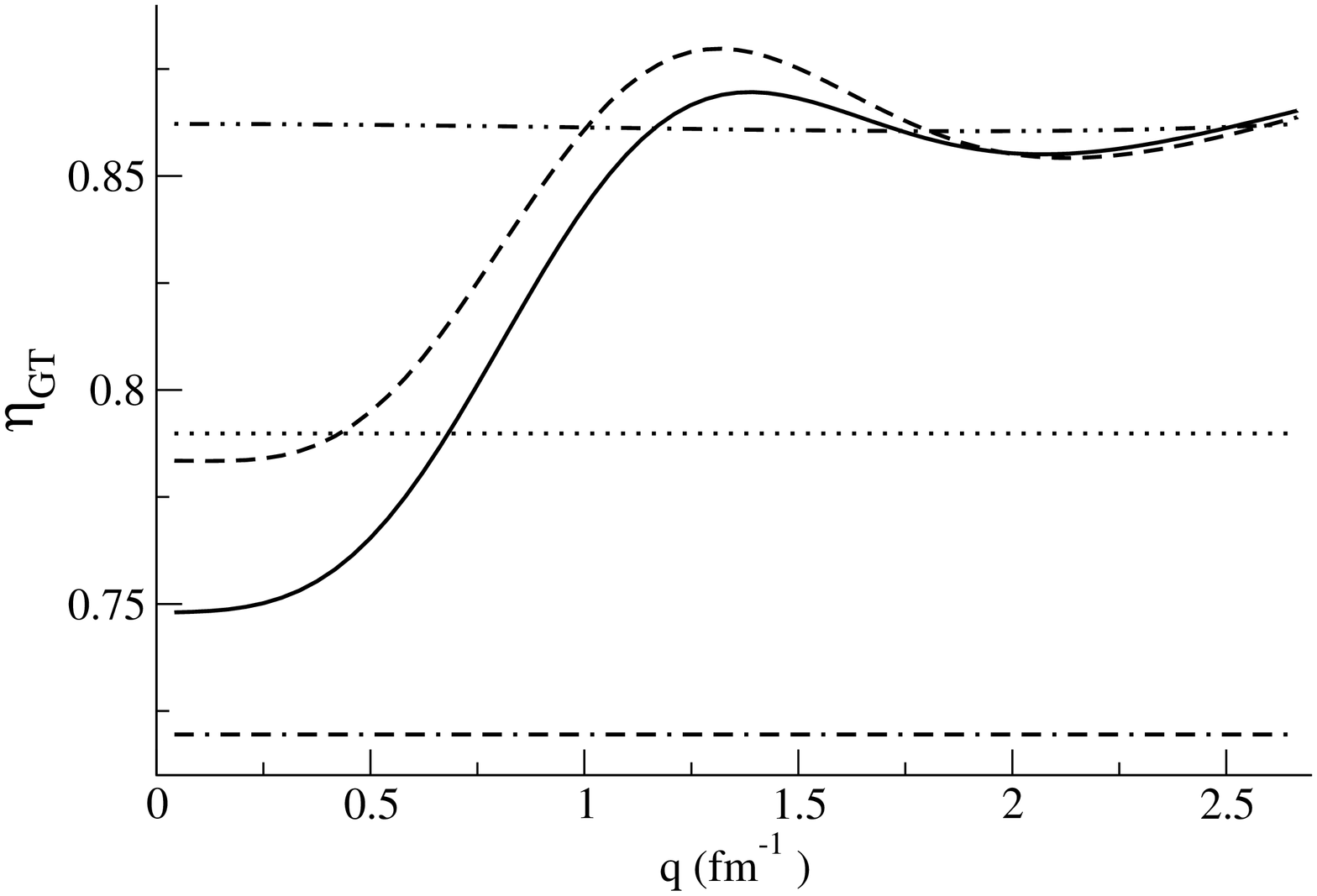}
\caption{Contributions to the $\gtav$ for for $k_N=k_{FN}$ and $\rho = \rho_0$:
the dash-double dot line includes all of the $q$-independent terms, while 
the solid line shows the full result.  See text for description of other 
curves.}
\label{gtparts}
\end{figure}

Fig. \ref{gtparts} illustrates the relative contributions of the various terms 
to $\gtav$.  As in the Fermi case, the main quenching comes from the $GT.2.d.a$ 
term; approximating the $|$CBME$|^2$ by $|1+GT.2.d.a |^2$ gives 
$\gtav = 0.79$ (dotted line).  It decreases to 0.72 on adding the $GT.n.e.a$ 
terms (double dash-dot line).  The double dot-dash line shows the result after 
including $GT.n.e.j$ terms which reduce the quenching.

The main $q$-dependence comes from the 1st order 
$GT.1.d.j$ term; results obtained after adding this term are shown by the 
dashed line.  The $GT.2.d.j$ term also contributes to the $q$-dependence 
(full line gives the total $\gtav$). 

The dash-dot and the double dot-dash lines have a barely visible 
$q$-dependence coming from the $GT.n.e.y$ terms.  In the Fermi case these 
exchange terms depend only on $k_n$ and $k_p$; however, in the GT case they 
introduce a dependence of $\gtav$ on 
the angle between ${\bf k}_n$ and ${\bf k}_p$.  This 
appears as a $q$-dependence, but it is very small ($< 0.002$). 

\begin{figure}[htbp] 
\includegraphics[height=15cm, angle=-90]{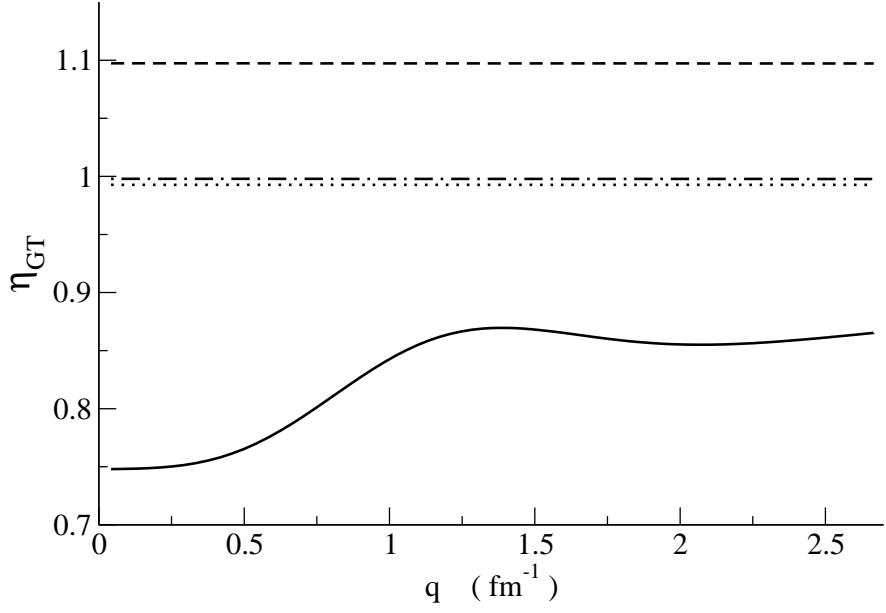}
\caption{Correlation dependence of $\gtav$ for for $k_N=k_{FN}$ and $\rho = \rho_0$.  The solid
line is for $\gtav$ with the full $F$.  The dashed line 
shows results with $\gfd=\gff=0$, and in addition, $f^c=1$ for the dash-dot 
line. The dotted line shows $\overline{|\sum_{n,x} GT.n.x.a|^2}$ when $\gfd=\gff=0$. } 
\label{gtnotau} 
\end{figure}

The relative contributions of various correlations to $\gtav$ are shown in 
Fig. \ref{gtnotau}.  The dashed and the dash-dot lines show results obtained 
after setting $\gfd=\gff=0$, and in addition $\gfa=1$ respectively.  The 
central correlations contribute mostly via the $GT.n.e.j$ terms; the dotted 
line close to $\gtav =1$ is obtained by setting $\gfd=\gff=0$ and including 
only the $GT.n.x.a$ terms. 

\begin{figure}[htbp] 
\includegraphics[height=15cm, angle=-90]{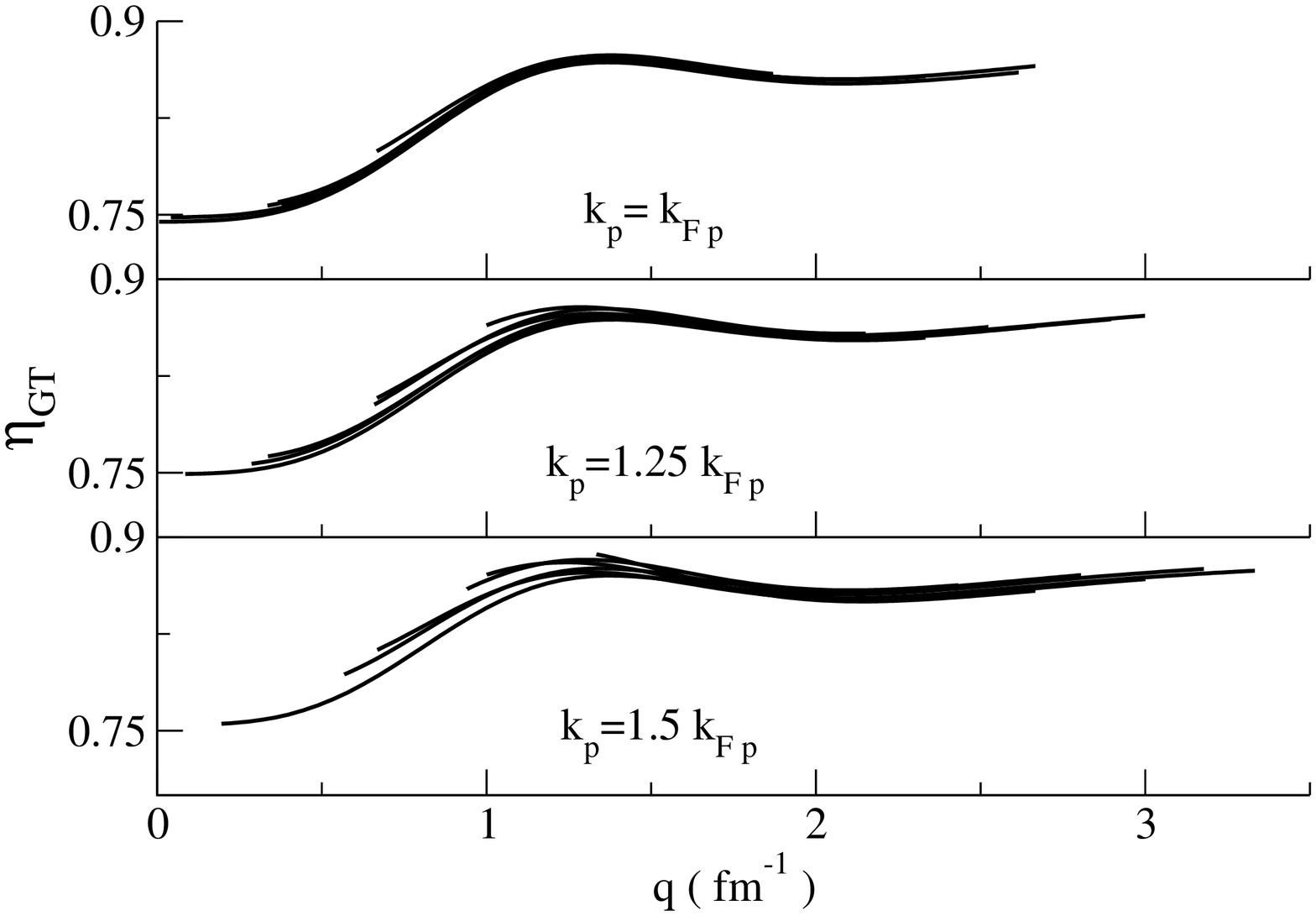}
\caption{Dependence of the $\gtav$ on the initial ($k_n$) 
and final ($k_p$) momenta for $\rho = \rho_0$.  
Each set contains six lines depicting the results for $k_n=(.5,~.75,~1)k_{Fn}$, 
and $x_p=0.3$ and 0.5 for the indicated value of $k_p$.} 
\label{knkpdepgt}
\end{figure}

The dependence of $\gtav$ on $k_n$ and $k_p$ is shown in Fig. \ref{knkpdepgt}. 
It is small $< 0.03$ as for $\eta_F$.

\section{Correlated Basis Neutral-Vector Matrix Element}

In ``one-body'' NV transitions the final state is:
\beq
\PhiF = a^{\dagger}_{{\bf k}_f \chi^f_{N}} a_{{\bf k}_i 
\chi^i_{N}} \PhiI ~,
\label{phif2}
\eeq
where $k_i \le k_{F N}$ and $k_f > k_{F N}$. 
The NV matrix element is nonzero
only when the initial and final spin-isospin states, 
$\chi_{N}^f$ and $\chi^i_{N}$ are the same.

The terms in the cluster expansion of the NV CBME are denoted by  
\beq
NV.n.x.y = -\sin^2\theta_W \ NV.n.x.y.1 + \frac{1}{2}(1-2\sin^2\theta_W)
\ NV.n.x.y.z ~,
\eeq
where $n, x$ and $y$ are defined in section \ref{CBFME} and $1$ and $z$
respectively represent contributions of $ e^{i {\bf q} \cdot {\bf r}_i}  $ and
$ \tau^z_i e^{i {\bf q} \cdot {\bf r}_i} $.
For the $NV.n.x.y.1$ terms we obtain: 
\ba
NV.0.d.a.1 & = & 1
\ea
\ba
NV.1.d.j.1 &=& \int \! \! d^3r \, e^{-i \mathbf{q}\cdot\mathbf{r}} 
\ 2\, \big((\gfa-1)\rho + \gfb (\rho_p-\rho_n) \tauaz \big) \\
NV.1.e.j.1 &=& - \frac{1}{2}\int \! \! d^3r \, e^{i \mathbf{k}_i\cdot\mathbf{r}} \Big\{ 
 (\gfa - 1+ 3 \gfc + 3\gfb + 9 \gfd) (\rho_p \, \ell_p + \rho_n \, \ell_n)
 \nonumber \\ 
 && + (\gfa - 1+ 3 \gfc - \gfb - 3 \gfd) (\rho_p \, \ell_p - \rho_n \, \ell_n)
  \tauaz \Big\}
\ea
\ba
NV.1.d.a.1 &=& 0 \\
NV.1.e.a.N.1 &=& - \frac{1}{2}\int \! \! d^3r \, e^{i \mathbf{k}_f\cdot\mathbf{r}} \Big\{ 
 (\gfa - 1+ 3 \gfc + 3\gfb + 9 \gfd) (\rho_p \, \ell_p + \rho_n \, \ell_n)
 \nonumber \\ 
 && + (\gfa - 1+ 3 \gfc - \gfb - 3 \gfd) (\rho_p \, \ell_p - \rho_n \, \ell_n)
  \tauaz \Big\} \\
NV.1.e.a.D.1 &=&   \frac{1}{4} \int \! \! d^3r \, 
(e^{i \mathbf{k}_i\cdot\mathbf{r}} +e^{i \mathbf{k}_f \cdot\mathbf{r}})
\big[ (\gfa - 1 + 3 \gfc + 3\gfb + 9 \gfd)(\rho_p \, \ell_p + \rho_n \, \ell_n)
\nonumber \\
&& + (\gfa - 1 + 3 \gfc - \gfb - 3 \gfd) (\rho_p \, \ell_p - \rho_n \, \ell_n)
 \tauaz \big]
\ea
\ba
NV.2.d.j.1&=& \int \! \! d^3r e^{-i \mathbf{q}\cdot\mathbf{r}}\Big[
(C^{00}_d + 3 C^{11}_d)\rho + 2(C^{10}_d - C^{11}_d) (\rho_p-\rho_n) \tauaz
\Big]\\
NV.2.e.j.1 &=& -\frac{1}{4} \int \! \! d^3 r \, e^{i \mathbf{k}_i\cdot\mathbf{r}}
\Big\{ (C^{00}_e + 6 C^{10}_e -3 C^{11}_e)(\rho_p \, \ell_p +\rho_n\, \ell_n) \nonumber \\
&& +(C^{00}_e - 2 C^{10}_e + 5 C^{11}_e)(\rho_p \, \ell_p-\rho_n \, \ell_n)\tauaz\Big\}
\ea
\ba
NV.2.d.a.1 &=& 0 \\
NV.2.e.a.N.1 &=& -\frac{1}{4} \int \! \! d^3 r \, e^{i \mathbf{k}_f\cdot\mathbf{r}}
\Big\{ (C^{00}_e + 6 C^{10}_e -3 C^{11}_e)(\rho_p \, \ell_p +\rho_n\, \ell_n) \nonumber \\
&& +(C^{00}_e - 2 C^{10}_e + 5 C^{11}_e)(\rho_p \, \ell_p-\rho_n \, \ell_n)\tauaz\Big\} \\
NV.2.e.a.D.1 &=&  \frac{1}{8} \int \! \! d^3r \, 
(e^{i \mathbf{k}_i\cdot\mathbf{r}} +e^{i \mathbf{k}_f \cdot\mathbf{r}})
\big[ (C^{00}_e + 6 C^{01}_e - 3 C^{11}_e)(\rho_p \, \ell_p + \rho_n \, \ell_n)
\nonumber \\
&& + (C^{00}_e - 2 C^{01}_e + 5 C^{11}_e) (\rho_p \, \ell_p - \rho_n \, \ell_n)
 \tauaz \big]
\ea
where $\tauaz = \langle \chi_N^f (a)| \tau^z_a|\chi_N^i (a)\rangle$.

The $NV.n.d.a.1$ terms are zero for $n>0$ because $e^{i{\bf q}\cdot{\bf r}_i}$ commutes with 
the static correlation operators.  Also note that the exchange, $NV.n.e.a.1$, terms are zero 
when $|{\bf k}_i|=|{\bf k}_f|$.

The $NV.n.x.y.z$ terms are given by: 
\ba 
NV.0.d.a.z & = & \tauaz
\ea
\ba
NV.1.d.j.z &=& \int \! \! d^3r \, e^{-i \mathbf{q}\cdot\mathbf{r}} 
\ 2\, \big((\gfa-1)(\rho_p-\rho_n) + \gfb \rho \tauaz \big) \\
NV.1.e.j.z &=& - \frac{1}{2}\int \! \! d^3r \, e^{i \mathbf{k}_i\cdot\mathbf{r}} \Big\{ 
 (\gfa - 1+ 3 \gfc + \gfb + 3 \gfd) (\rho_p \, \ell_p - \rho_n \, \ell_n)
 \nonumber \\ 
 && + (\gfa - 1+ 3 \gfc + \gfb + 3 \gfd) (\rho_p \, \ell_p + \rho_n \, \ell_n)
  \tauaz \Big\}
\ea
\ba
NV.1.d.a.z &=& 0 \\
NV.1.e.a.N.z &=& - \frac{1}{2}\int \! \! d^3r \, e^{i \mathbf{k}_f\cdot\mathbf{r}} \Big\{ 
 (\gfa - 1+ 3 \gfc + \gfb + 3 \gfd) (\rho_p \, \ell_p - \rho_n \, \ell_n)
 \nonumber \\ 
 && + (\gfa - 1+ 3 \gfc + \gfb + 3 \gfd) (\rho_p \, \ell_p + \rho_n \, \ell_n)
  \tauaz \Big\}\\
NV.1.e.a.D.z &=&  \tauaz  NV.1.e.a.D.1
\ea
\ba
NV.2.d.j.z&=& \int \! \! d^3r e^{-i \mathbf{q}\cdot\mathbf{r}}\Big[
(C^{00}_d - C^{11}_d)(\rho_p-\rho_n) + 2(C^{10}_d + C^{11}_d) \rho \tauaz
\Big]\\
NV.2.e.j.z &=& -\frac{1}{4} \int \! \! d^3 r \, e^{i \mathbf{k}_i\cdot\mathbf{r}}
\Big\{ (C^{00}_e + 2 C^{10}_e + C^{11}_e)(\rho_p \, \ell_p-\rho_n\, \ell_n) \nonumber \\
&& +(C^{00}_e + 2 C^{10}_e + C^{11}_e)(\rho_p \, \ell_p+\rho_n \, \ell_n)\tauaz\Big\}
\ea
\ba
NV.2.d.a.z &=& \int \! \! d^3 r 
\big( 4 C^{11}_d (\rho_p-\rho_n) - 4 C^{11}_d \rho \tauaz \big) \\
NV.2.e.a.N.z &=& -\frac{1}{4} \int \! \! d^3 r \, e^{i \mathbf{k}_f\cdot\mathbf{r}}
\Big\{ (C^{00}_e + 2 C^{10}_e + C^{11}_e)(\rho_p \, \ell_p - \rho_n \, \ell_n)
\nonumber \\
&& + (C^{00}_e + 2 C^{10}_e + C^{11}_e)(\rho_p \, \ell_p + \rho_n \, \ell_n)
\tauaz \Big\} \\
NV.2.e.a.D.z &=& \tauaz NV.2.e.a.D.1
\ea
In symmetric nuclear matter the matrix elements of $\tau_z$ are related to those 
of $\tau^{\pm}$.  In this case the $NV.n.x.y.z = F.n.x.y$.  However, when $x_p <
0.5$ the $NV$ matrix 
elements have additional terms dependent upon $\rho_n-\rho_p$, or equivalently $x_p$.  

\subsection{Results of Neutral Vector Matrix Element}

\begin{figure}[htbp]
\includegraphics[height=15cm, angle=-90]{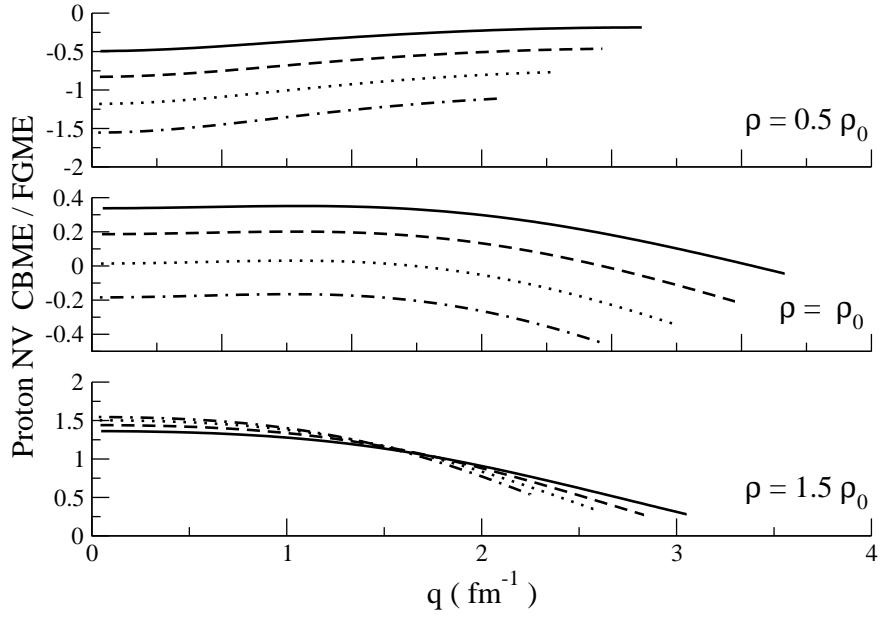}
\caption{Proton NV CBME scaled by FGME as a function of $q$ and proton fraction
$x_p$ for $k_i=k_f=k_{F p}$.  The solid, dashed, dotted and dash-dot lines 
show results for $x_p=0.5,~0.4,~0.3,$ and 0.2.}
\label{NVpFull}
\end{figure}

In uncorrelated FG states, the neutral vector matrix element is:
\beq
-\sin^2 \theta_W + 
\frac{1}{2}(1-2 \sin^2 \theta_W) \tauaz = 
-0.2314 \pm 0.2686
\label{e.nvfg}
\eeq
for proton and neutron particle-hole pairs respectively.  The above two terms nearly 
cancel for uncorrelated protons.  The correlations influence each operator differently 
and the final CB result depends sensitively on 
$k_i, k_f, \rho$ and $x_p$.  The strong dependence of the proton NV matrix
element on $\rho$ and $x_p$ is shown 
in Fig. \ref{NVpFull} where we have
plotted the proton particle-hole NV CBME scaled by 0.0372, the FGME.
Note that the value of the CBME (not $|CBME|^2$) is shown in this figure. 
At low densities the first term dominates, and the CBME is negative; 
however, at higher densities the second term becomes larger, and the matrix
element becomes positive. 
At $\rho \sim \rho_0$ the cancellation of the two terms is almost exact, and the 
proton NV CBME is very small.  Fortunately, in this case the FGME is small and
the CBME is of the same order in the considered density range.  Thus, the
coupling of the proton NV current is not likely to have a significant
contribution to the $\nu$-nucleus interaction.

\begin{figure}[htbp]
\includegraphics[height=15cm, angle=-90]{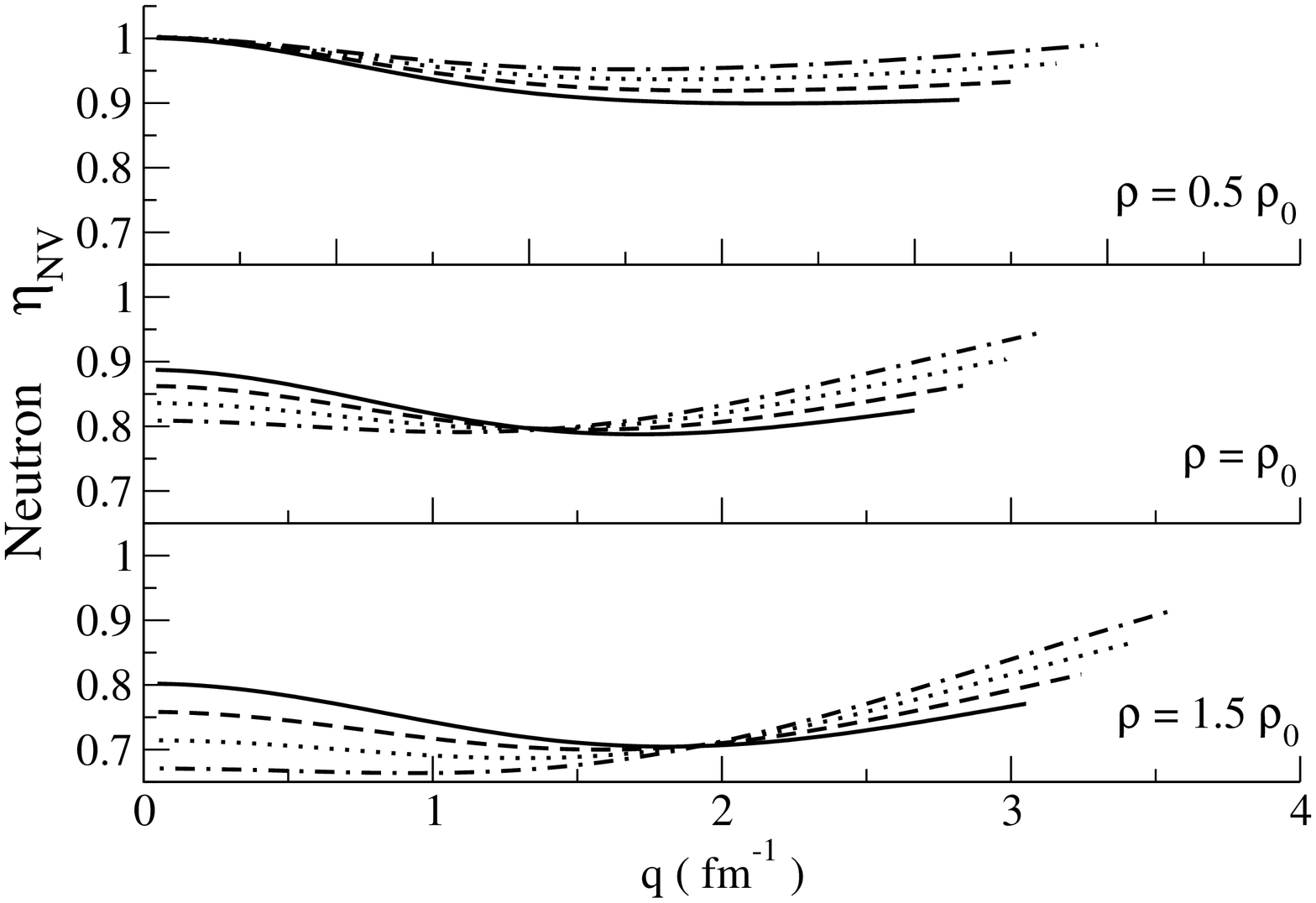}
\caption{Nuetron $\eta_{NV}$ as a function of $q$ and proton fraction
$x_p$ for $k_i=k_f=k_{F n}$.  The solid, dashed, dotted and dash-dot lines 
show results for $x_p=0.5,~0.4,~0.3,$ and 0.2.}
\label{NVnFull}
\end{figure}

Figure \ref{NVnFull} shows the density and $x_p$ dependence of $\eta_{NV}$ 
for neutron particle-hole pair excitations.
At $\rho = \frac{1}{2} \rho_0$ the correlations increase the contribution of the 
first term and decrease that of the second term in Eq. (\ref{e.nvfg}) by a similar 
magnitude.  Therefore at small $\rho$ and $q$ the NV neutron CBME $\sim$ FGME. 
However, at higher densities it is quenched. As mentioned earlier these matrix elements 
have a significant $x_p$ dependence absent in the charge current matrix elements. 

\begin{figure}[htbp] 
\includegraphics[height=15cm, angle=-90]{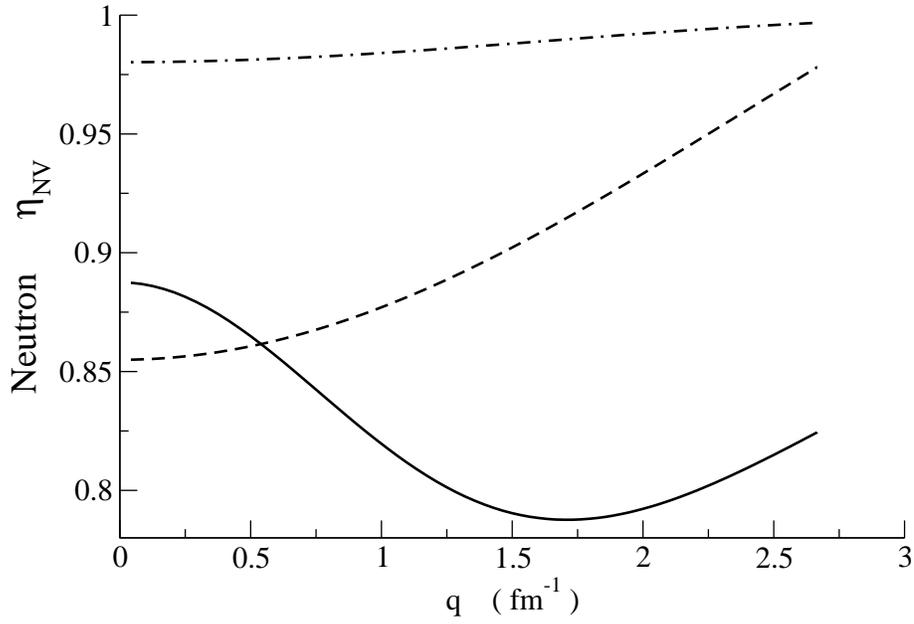}
\caption{Correlation dependence of the neutron $\eta_{NV}$ for 
$k_i=k_f=k_{F n}$ and $\rho = \rho_0$. 
  The dashed line  shows results with $\gfd=\gff=0$,
and in addition, $f^c=1$ for the dash-dot  line.
The solid line gives the full result.} 
\label{nvcorparts} 
\end{figure}

Fig. \ref{nvcorparts} shows the contributions of the various correlations  to the
NV neutron CBME.  The CBME is influenced by contributions of the 
$f^c-1$  correlations to $NV.n.x.j.1$ and those of the 
$\gfd (r_{ij})\sigij \tauij$ and $\gff (r_{ij})S_{ij} \tauij$ correlations to 
$NV.n.x.y.z$ terms.  The results obtained after setting $\gfd = \gff = 0$, and 
in addition $\gfa =1$ are shown by dashed and dash-dot lines in Fig. \ref{nvcorparts}. 

The Neutral Vector CBME for a neutron particle-hole pair 
does not depend significantly on the magnitudes
of the initial and final nucleon  momenta.
Variation of $k_i$ from 0.5 to 1 and of $k_f$ from 1 to 1.5 $k_{F n}$ changes 
$\eta_{NV}$ by less than 3 \%. 

\section{Correlated Basis Neutral-Axial-Vector Matrix Element} 

The operator,
$\mathbf{O}_{NA}$ is an axial vector and it is convenient to express its matrix
element using the following two axial vectors, similar to those used for the
Gamow-Teller CBME (Sect. III):
\beq
\sigab = \langle \chi^f_{N}(a) | \, \boldsigma(a) | \chi^i_{N}(a) \rangle
\eeq
and
\beq
\Atb = 3 \ \hat{\bf r}_{aj} \ \sigab \cdot \hat{\bf r}_{aj} - \sigab ~ .
\eeq
We assume that $\chi_N^i$ in
Eq. (\ref{phif2}) is spin up  and calculate the sum of the square of the NA 
matrix element for  the two
final states with $\chi_N^f = \uparrow , \downarrow$
for both $N=n$ and $p$.
The terms in the cluster expansion of the NA CBME are denoted by  $NA.n.x.y$
as in Section \ref{CBFME},  and the factor $g_A$ is omitted 
for brevity.  We obtain:  
\ba 
NA.0.d.a & = & \frac{1}{2}\sigab \tauaz
\ea
\ba
NA.1.d.j &=& \frac{1}{2}\int \! \! d^3r \, e^{-i \mathbf{q}\cdot\mathbf{r}} 
\ 2\, \big[ (\rho_p-\rho_n)(\gfc \sigab  + \gfe \Atb) \nonumber \\
&& + \rho \tauaz( \gfd \, \sigab 
+ \gff \, \Atb )\big] \\
NA.1.e.j &=& - \frac{1}{4}\int \! \! d^3r \, e^{i \mathbf{k}_i\cdot\mathbf{r}} \Big\{ 
((\gfa - 1 + \gfc + \gfb - 3 \gfd) \sigab + (\gfe + 3 \gff) \Atb )(\rho_p\, \ell_p
-\rho_n\, \ell_n) \nonumber \\
&&+((\gfa - 1 + \gfc + \gfb + 5 \gfd) \sigab + (\gfe - \gff) \Atb )(\rho_p\, \ell_p
+\rho_n\, \ell_n)\tauaz \big\}
\ea
\ba
NA.1.d.a &=& 0 \\
NA.1.e.a.N &=&- \frac{1}{4}\int \! \! d^3r \, e^{i \mathbf{k}_f\cdot\mathbf{r}} \Big\{ 
((\gfa - 1 + \gfc + \gfb - 3 \gfd) \sigab + (\gfe + 3 \gff) \Atb )(\rho_p\, \ell_p
-\rho_n\, \ell_n) \nonumber \\
&&+((\gfa - 1 + \gfc + \gfb + 5 \gfd) \sigab + (\gfe - \gff) \Atb )(\rho_p\, \ell_p
+\rho_n\, \ell_n)\tauaz \big\} \\
NA.1.e.a.D &=&  \frac{1}{2}\sigab \tauaz  NV.1.e.a.D.1 
\ea
\ba
NA.2.d.j&=& \frac{1}{2}\int \! \! d^3r e^{-i \mathbf{q}\cdot\mathbf{r}}\Big[
\big(\gfoojdc-\gfTTjdc \big) \,\sigab +\big(\gfoojdt -\gfTTjdt \big) \, \Atb
\Big] (\rho_p-\rho_n) \nonumber \\
&&+ \big(\gfTojdc + \gfoTjdc +2\gfTTjdc \big) \,\sigab +
\big(\gfTojdt + \gfoTjdt +2\gfTTjdt \big) \, \Atb
\Big] \rho \tauaz \\
NA.2.e.j &=& -\frac{1}{8} \int \! \! d^3 r \, e^{i \mathbf{k}_i\cdot\mathbf{r}}
\Big\{\big( \gfoojec + 3 \gfTojec - \gfoTjec + \gfTTjec\big) \sigab \nonumber \\
&& + \big( \gfoojet + 3 \gfTojet - \gfoTjet + \gfTTjet\big) \Atb \Big\} (\rho_p\,
\ell_p - \rho_n \, \ell_n) \nonumber \\
&&+\Big\{\big( \gfoojec - \gfTojec +3 \gfoTjec + \gfTTjec\big) \sigab \nonumber\\
&& + 
\big( \gfoojet - \gfTojet +3 \gfoTjet + \gfTTjet\big) \Atb \Big\} (\rho_p\,
\ell_p + \rho_n \, \ell_n) \tauaz
\ea
\ba
NA.2.d.a &=& \frac{1}{2}\int \! \! d^3r \Big[
\big(\gfTojdc + \gfoTjdc +2\gfTTjdc -2(C^{01}_d - C^{11}_d)\big) \,\sigab
(\rho_p-\rho_n) \nonumber \\
&&+ \big(\gfoojdc-\gfTTjdc - C^{00}_d - 3 C^{11}_d \big) \,\sigab 
  \rho \tauaz \Big]\\
NA.2.e.a.N&=& -\frac{1}{8} \int \! \! d^3 r \, e^{i \mathbf{k}_f\cdot\mathbf{r}}
\Big\{\big( \gfoojec - \gfTojec +3 \gfoTjec + \gfTTjec\big) \sigab \nonumber \\
&& + \big( \gfoojet - \gfTojet +3 \gfoTjet + \gfTTjet\big) \Atb \Big\} (\rho_p\,
\ell_p - \rho_n \, \ell_n) \nonumber \\
&&+\Big\{\big( \gfoojec +3 \gfTojec - \gfoTjec + \gfTTjec\big) \sigab \nonumber\\
&& + 
\big( \gfoojet +3 \gfTojet - \gfoTjet + \gfTTjet\big) \Atb \Big\} (\rho_p\,
\ell_p + \rho_n \, \ell_n) \tauaz \\
NA.2.e.a.D &=&  \frac{1}{2}\sigab \tauaz NV.2.e.a.D.1
\ea

\subsection{Results of Neutral Axial-Vector Matrix Element} 

\begin{figure}[htbp]
\includegraphics[height=15cm, angle=-90]{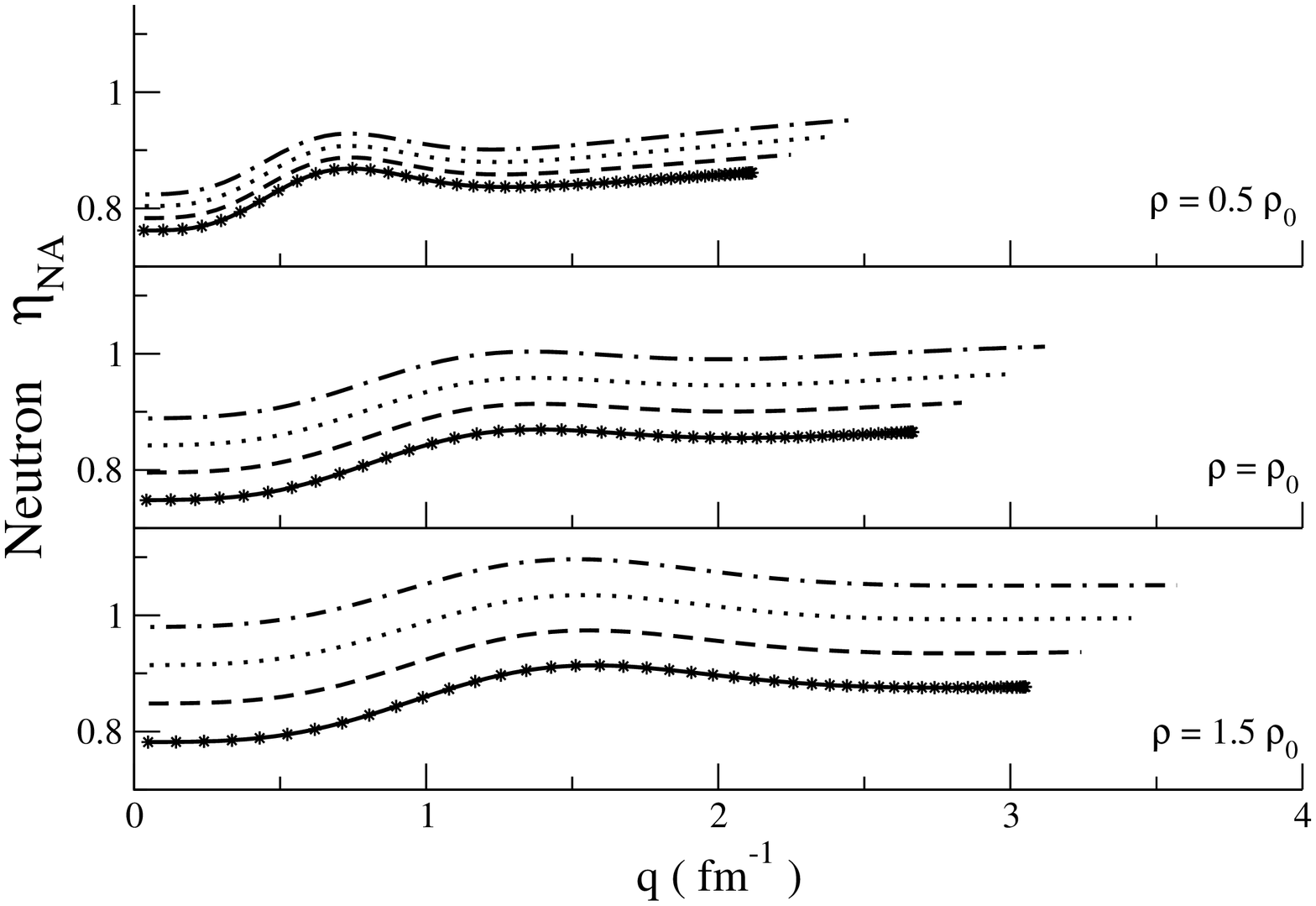}
\caption{Neutron $\eta_{NA}$ as a function of $q$ and proton fraction
$x_p$ for $k_i=k_f=k_{F n}$.  The solid, dashed, dotted and dash-dot lines 
show results for $x_p=0.5,~0.4,~0.3,$ and 0.2.  The stars are results for $\gtav$
at $x_p = 0.5$.}
\label{NAFulln}
\end{figure}

\begin{figure}[htbp]
\includegraphics[height=15cm, angle=-90]{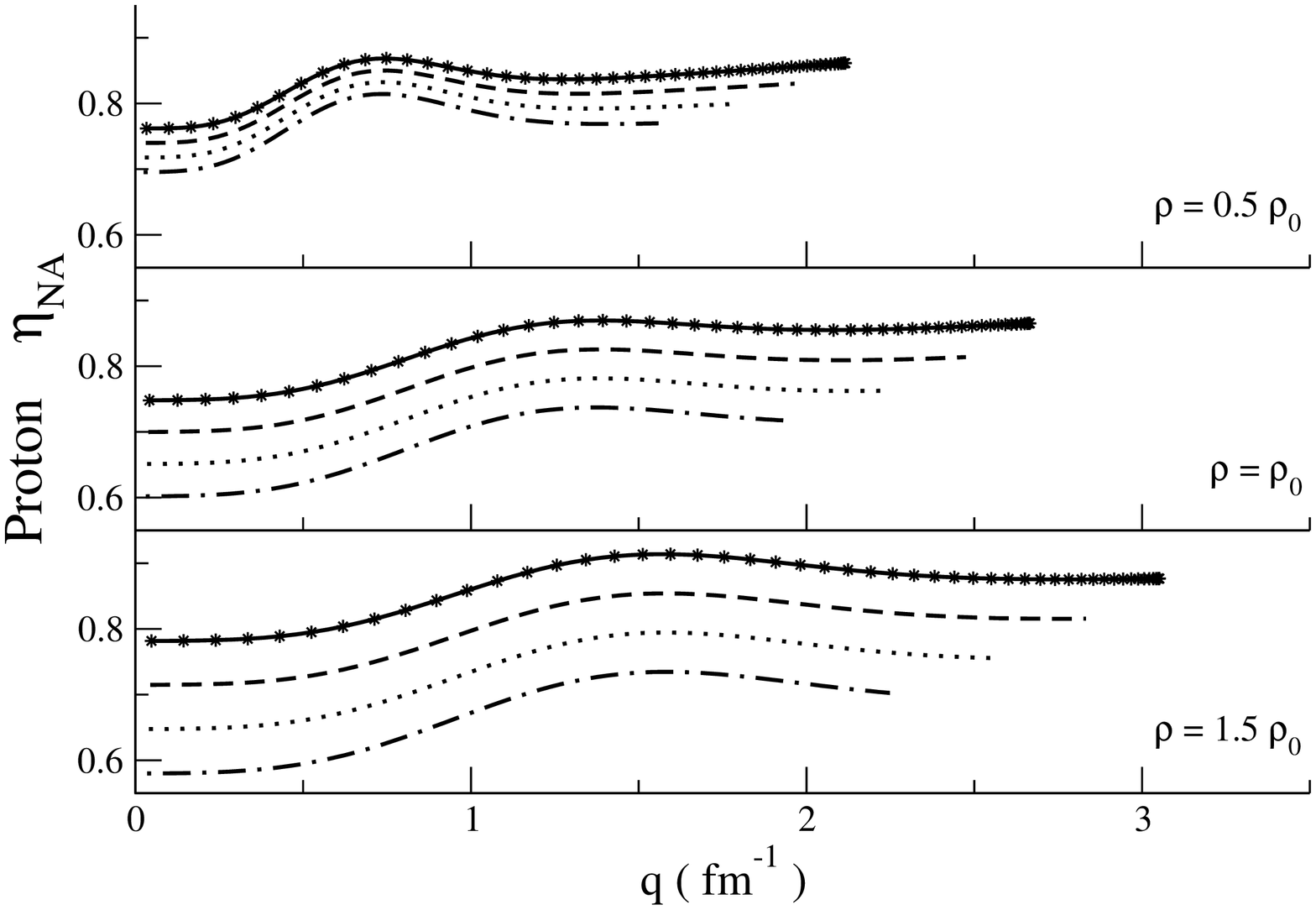}
\caption{Proton $\eta_{NA}$ as a function of $q$ and proton fraction
$x_p$ for $k_i=k_f=k_{F p}$.  The solid, dashed, dotted and dash-dot lines 
show results for $x_p=0.5,~0.4,~0.3,$ and 0.2.  The stars are results for $\gtav$
at $x_p = 0.5$.}
\label{NAFullp}
\end{figure}

We discuss only the sum of the $|$CBME$|^2$ over the two final spin states 
because it is independent  of the chosen spin quantization 
axis.  This sum equals 3/4 for FGME.  In the following we provide results for: 
\beq 
\eta_{NA}
\equiv \frac{4}{3}(|\langle F \uparrow |\mathbf{O}_{NA}| I \rangle|^2+ |\langle
F \downarrow |\mathbf{O}_{NA}| I \rangle|^2) 
\eeq
The $\av$ for neutron and proton particle-hole pairs are
plotted in Figures \ref{NAFulln} and \ref{NAFullp} respectively for 
the considered density and proton fraction values.  In these 
matrix elements $k_i=k_f=k_{FN}$. 

The charge-changing and neutral axial vector operators (${\bf O}_{GT}$ and
${\bf O}_{NA}$), appropriately scaled, can be
interpreted as the three components of an isospin vector 
operator.  In symmetric nuclear matter the expectation values of these three components 
are equal as one can not quantify the isospin axis.  The stars in
Figures \ref{NAFulln} and \ref{NAFullp} are results obtained for
$\gtav$ for symmetric nuclear matter with equivalent initial and final momenta
and densities.  They are identical to those obtained for $\av$ for both proton
and neutron particle-hole pairs.

Unlike the results for the GT CBME, there is a noticeable dependence of $\av$
on the proton fraction at all densities considered.  This ($\rho_p - \rho_n$) 
dependence originates from the $\tau^z_j$ in $NA.n.x.j$ and 
$NA.n.x.a$ terms.  We can approximate the NA
results obtained for $x_p < 0.5$ by adding a density dependent term
proportional to  ($\rho_p - \rho_n$) to $\av$ for symmetric nuclear
matter.  For small $q$ this approximation is:
\ba
\eta_{NA}(\rho, x_p < 0.5)  &=&
\eta_{NA}(\rho, x_p = 0.5) - C_{N}(\rho) (\rho_p -
\rho_n) \\
&=& \eta_{GT}(q=0) + \alpha_{GT} q^2 - C_{N}(\rho) (\rho_p -
\rho_n)~,
\ea
where we have used $\av = \gtav$ at $x_p = 0.5$ and Eq. (\ref{quad}).  The
values obtained for $C_N (\rho)$ at the three densities considered are given in
Table \ref{table3}.

The correlation dependence and initial and final momenta dependence studied for
$\gtav$ are applicable here and will not be discussed further.

\section{Correlated Basis Interaction}
\label{CBI}

The expectation value of $H-T_{FG}(X)$, where $T_{FG}(X)$ is the kinetic energy
of the Fermi-gas state $\Phi_X$,
is expanded to calculate the energy of the correlated 
state $|X\rangle$.  It is given by: 
\ba
\langle X| H |X \rangle  
&=& \frac{\langle \Phi_X|[{\cal S} \Pi F_{ij}] \ (H-T_{FG}(X))\ [{\cal S} \Pi F_{ij}]| \Phi_X \rangle}
{ \langle \Phi_X|[{\cal S} \Pi F_{ij}]^2| \Phi_X \rangle} + T_{FG}(X)~, \\
T_{FG}(X) &=& \sum_{all~i~occupied~in~\Phi_X} \frac{k_i^2}{2m}~. 
\ea
Since $\Phi_X$ is an eigenstate of the kinetic energy operator $T=\sum_i
-\nabla^2_i/2m$, with 
eigenvalue $T_{FG}(X)$, it is not necessary to expand the FG kinetic energy.  The 
$(H-T_{FG}(X))|X\rangle$ does not contain terms with $\nabla_i^2$ operating on $| \Phi_X \rangle$. 
Including only two-body clusters we obtain:
\beq
\langle X| H |X \rangle =  T_{FG}(X) + \sum_{i<j} 
\langle ij-ji|F_{ij}\Big[ v_{ij}F_{ij} -\frac{1}{m}(\nabla^2F_{ij}) -\frac{2}{m} 
(\boldnabla F_{ij}) \cdot \boldnabla \Big]|ij \rangle ~, 
\label{e2b}
\eeq
where $|ij \rangle = e^{i({\bf k}_i \cdot {\bf r}_i + {\bf k}_j \cdot {\bf r}_j)}
\chits(i) \chits(j)$.
The gradient operates on the relative coordinate, 
and the sum $i<j$ is over states occupied in $\Phi_X$.  The effective 
correlated basis two-nucleon interaction (CBI) is given by (see Eq. (8)):
\beq
v^{CB}_{ij} = F_{ij}\Big[ v_{ij}F_{ij} -\frac{1}{m}(\nabla^2F_{ij}) -\frac{2}{m}
(\boldnabla F_{ij}) \cdot \boldnabla \Big]  
\eeq
in the 2-body cluster approximation. The energies of correlated states $|X\rangle$ are obtained 
by using this $v^{CB}_{ij}$ in 1st order with FG wave functions, $\Phi_X$, as in the
Hartree-Fock approximation. 

The $v_{ij}^{CB}$ has a momentum dependence via the $(\boldnabla F_{ij}) \cdot \boldnabla$ 
term which gives contributions to the matter energy via exchange terms in Eq.
(\ref{e2b}).  
This contribution is much smaller than that of the momentum independent, static terms in $v_{ij}^{CB}$   
defined as: 
\beq
v^{CBS}_{ij} = F_{ij} ( v_{ij} - \frac{1}{m} \nabla^2 ) F_{ij}~. 
\label{vcbs} 
\eeq

In the present work we have considered only the static part of $F_{ij}$ as mentioned 
in the introduction.  We therefore keep only the dominant, 
static part of the full Argonne $v_{ij}$.  The full $v_{ij}$ is first approximated by 
a $v_8^{\prime}$ interaction chosen such that it equals the 
isoscalar part of the full interaction in all $S$ and $P$ waves as well as 
in the $^3D_1$ wave and its coupling to the $^3S_1$.  The difference between the 
full and the $v_8^{\prime}$ interactions is small and treated perturbatively in 
the quantum Monte Carlo calculations \cite{PPCPW97}.  The $v_8^{\prime}$ has 
terms with the six static operators, $O^{p=1,6}_{ij}$,  and two spin-orbit terms. 
The later two are omitted to obtain the static part of Argonne $v_{ij}$. 
In this approximation the $v^{CBS}$ is a static operator having six terms with $O^{p=1,6}$: 
\beq
v^{CBS}_{ij} = \sum_{p=1,6} v_p^{CBS}(r_{ij}) O^p_{ij} ~. 
\label{vcbsip} 
\eeq

The Landau-Migdal effective interactions used in studies of weak interactions in 
nuclei \cite{KLV99} and nucleon matter \cite{PLSV01} are obtained from the spin-isospin 
susceptibilities of nucleon matter.  We have therefore studied these susceptibilities 
with the $v^{CB}$ and $v^{CBS}$.  The energy of nucleon matter 
with densities $\rho_{N \uparrow}$ and $\rho_{N \downarrow}$ can be expressed as:
\ba
\label{exyz}
E(\rho,x,y,z) &=& E_0(\rho) +E_{\tau}(\rho) x^2 + E_{\sigma}(\rho) y^2 + E_{\sigma \tau} (\rho) z^2 ~, \\
x &=& (\rho_{n \uparrow} + \rho_{n \downarrow} - \rho_{p \uparrow} - \rho_{p \downarrow})/ \rho ~, \\ 
y &=& (\rho_{n \uparrow} - \rho_{n \downarrow} + \rho_{p \uparrow} - \rho_{p \downarrow})/ \rho ~, \\ 
z &=& (\rho_{n \uparrow} - \rho_{n \downarrow} - \rho_{p \uparrow} + \rho_{p \downarrow})/ \rho ~.  
\ea
The $\tau, \sigma$ and $\sigma \tau$ susceptibilities are proportional to 
$E_{\tau, \sigma, \sigma \tau}^{-1}$, and $E_0(\rho)$ is the energy of symmetric nuclear 
matter with $x=y=z=0$.  Note that the $E_{\tau}(\rho_0)$ is the 
familiar symmetry energy in the liquid drop mass formula.  In principle, the above 
expansion is valid at small values of $x,y,$ and $z$; however, within the accuracy of 
available calculations it seems to be valid up to $x=1$ \cite{LP81, BL91}.  

\begin{figure}[htbp]
\includegraphics[height=15cm, angle=-90]{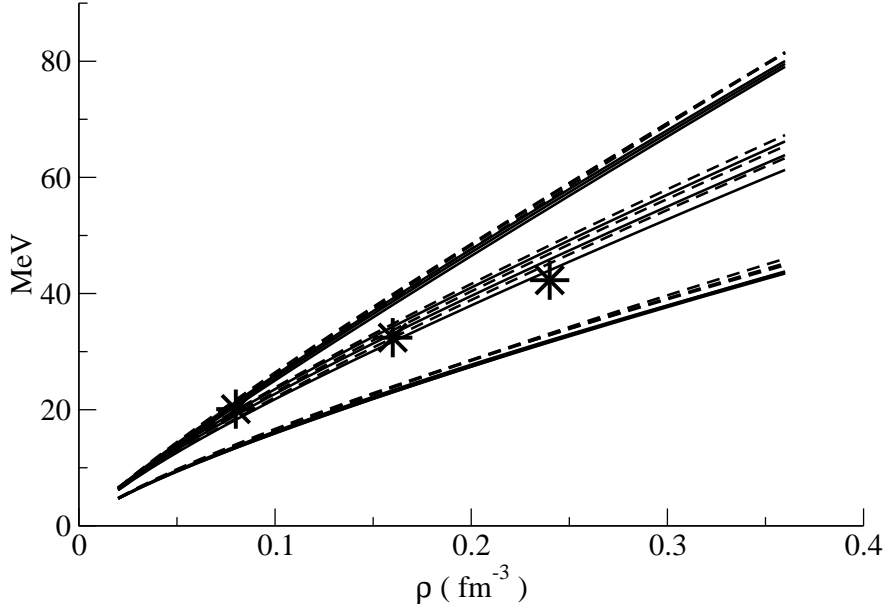}
\caption{$E_{\sigma \tau}(\rho)$ (upper set), $E_{\tau}(\rho)$ (middle set) and
$E_{\sigma}(\rho)$ (lower set) of symmetric nuclear matter.  In each set, the upper most curves are results using $F_{ij}$ for $\rho =
\frac{1}{2} \rho_0$, the middle for $\rho = \rho_0$, and the lowest for $\rho =
\frac{3}{2} \rho_0$. Solid lines show the results for $v^{CB}$ and the dashed
lines $v^{CBS}$.  Stars denote values obtained for $E_{\tau} (\rho)$ from
variational calculations \cite{MPR02}.}
\label{snms}
\end{figure}

We have calculated the $E_{\tau, \sigma, \sigma \tau}(\rho)$ using the $v^{CB}$ obtained 
from the $F_{ij}$ at $\rho = \frac{1}{2},~1$ and $\frac{3}{2} \, \rho_0$.  
The results obtained with the 
$v^{CB}$ are given by full lines in Fig. \ref{snms}, while those with the simpler $v^{CBS}$ 
by dashed lines.  The momentum dependent part of $v^{CB}$ gives rather small contributions 
which may be neglected in the first approximation.  The $v^{CB}$ has a density dependence 
due to that of $F_{ij}$.  However, it has very little effect on $E_{\sigma}$ and $E_{\sigma \tau}$; 
the results obtained from the $\frac{1}{2}~, 1~,\frac{3}{2}~\rho_0$ $v^{CB}$'s essentially overlap. 
The density dependence of $v^{CB}$ has a small but noticeable effect on the symmetry 
energy $E_{\tau}(\rho)$.  

The stars on Fig. \ref{snms} show the values of $E_{\tau}(\rho)$ 
extracted from recent variational calculations \cite{MPR02} of symmetric nuclear matter (SNM) and 
pure neutron matter (PNM) with the Argonne 
v18 and Urbana IX interactions, assuming that Eq. (\ref{exyz}) is valid up to $x=1$ for $y=z=0$.  
The two-body $v^{CB}$ seems to provide a fair approximation to the $E_{\tau}$.  

\begin{figure}[htbp]
\includegraphics[height=15cm, angle=-90]{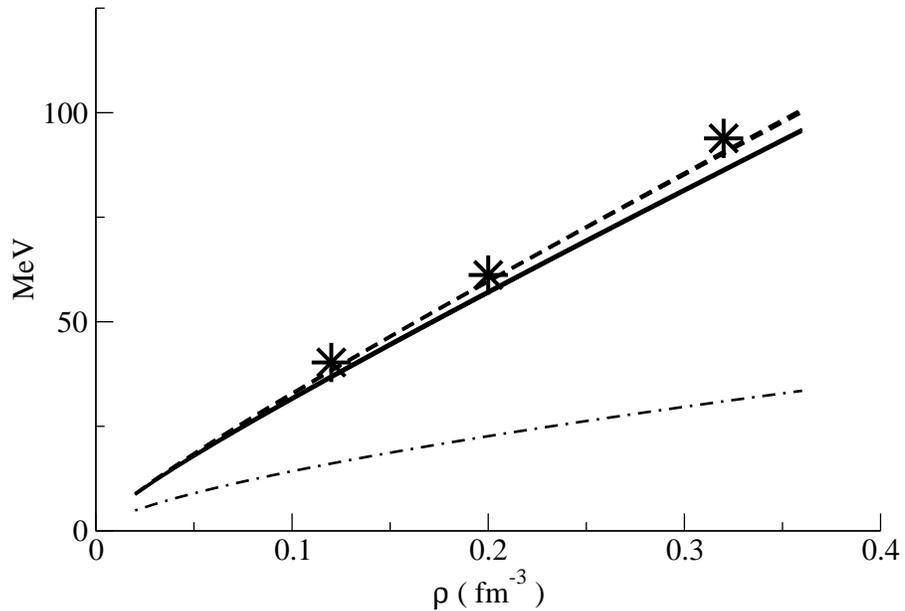}
\caption{$E_{\sigma}(\rho)$ for pure neutron matter.  The solid line shows
results obtained using $v^{CB}$ and the dashed for the 
$v^{CBS}$.  The results obtained with $F_{ij}$ for $\rho = \frac{1}{2},
1, \frac{3}{2} \rho_0$ are essentially indistinguishable. 
Stars denote values obtained for $E_{\sigma}^{PNM}(\rho)$ from quantum Monte
Carlo calculations \cite{FSS01}.  The dash-dot line is the Fermi-gas $E_{\sigma}(\rho)$.}
\label{pnms}
\end{figure}

We also consider the spin susceptibility of PNM given by the inverse 
of $E_{\sigma}^{PNM}(\rho)$ defined as:
\beq
E^{PNM}(\rho,y) = E^{PNM}_0(\rho) + E^{PNM}_{\sigma}(\rho) y^2 ~. 
\eeq
The results obtained with the $v^{CB}$ and $v^{CBS}$ are shown in
Fig. \ref{pnms} along with those obtained from quantum Monte Carlo calculations
\cite{FSS01} with the  static part of Argonne v18 and Urbana-IX interactions. 
The two-body $v^{CB}$ using $F_{ij}$ of SNM gives fairly accurate values of $E^{PNM}_{\sigma}$.

\begin{figure}[htbp]
\includegraphics[height=15cm, angle=-90]{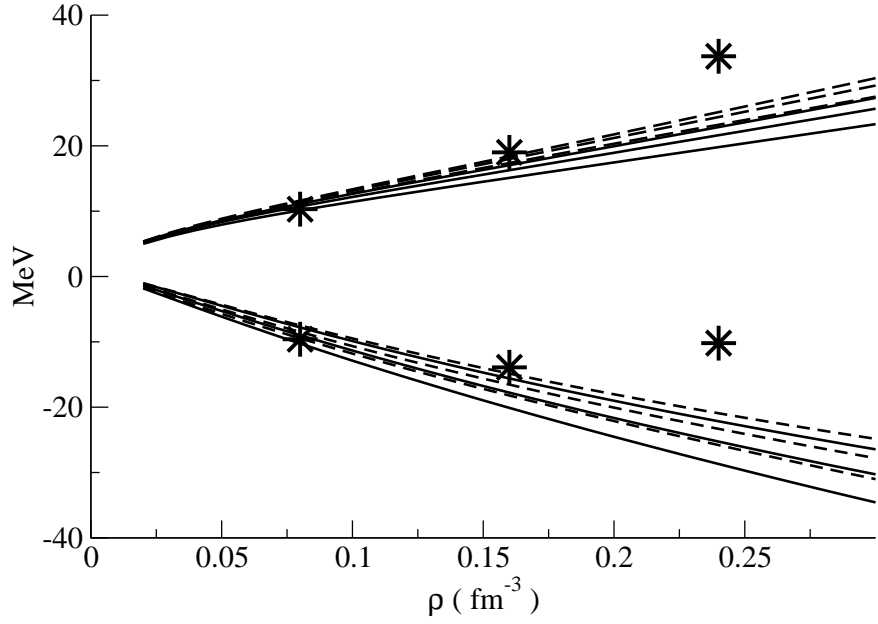}
\caption{$E_0(\rho)$ for symmetric nuclear matter (lower set of curves) and
pure nuetron matter (upper set of curves).
In each set, the upper most curves are results using $F_{ij}$ for $\rho =
\frac{3}{2} \rho_0$, the middle for $\rho = \rho_0$, and the lowest for $\rho =
\frac{1}{2} \rho_0$. Solid lines show the results for $v^{CB}$ and the dashed
lines $v^{CBS}$.  Stars denote values obtained for $E_{0} (\rho)$ from
variational calculations \cite{MPR02}.}
\label{gse}
\end{figure}

Fig. \ref{gse} shows $E_0(\rho)$ and $E_0^{PNM}(\rho)$ calculated from the $v^{CB}$ at 
the three values of $\rho$.  The stars in this figure give results of the recent 
variational calculations \cite{MPR02} with the full Argonne v18 and Urbana IX 
interactions.  At low densities the two-body $v^{CB}$ is not a bad approximation; however, 
the $E_0(\rho)$ obtained from it does not show a minimum at $\rho_0$.  The 3-body 
interaction and cluster contributions are repulsive and are essential to obtain the 
minimum.  

The 2-body $v^{CB}$ is more accurate in predicting the susceptibilities than the 
equation of state, $E_0(\rho)$.  This is partly because the contributions of $T_{FG}$ 
and $v^{CB}$ to the $E_{\tau, \sigma, \sigma \tau}(\rho)$ add.  The contribution of 
$T_{FG}$ to the $E^{PNM}_{\sigma}$ is shown in Fig. \ref{pnms}, it is about half 
of the total.  For this reason even relatively simple estimates \cite{old} 
of $E^{PNM}_{\sigma}$ are not too different from the current state of the art
\cite{FSS01}.  In contrast, in SNM 
the large negative $\langle v^{CB} \rangle$ cancels the $T_{FG}$ to produce 
a relatively small binding energy.  Therefore the many-body clusters 
are relatively more important in the calculation of $E_0(\rho)$.  

The results of the recent SNM calculations, which provided the $F_{ij}$ used here, 
are summarized in Table \ref{t:snm}.  The 1- and 2-body cluster contributions are 
calculated exactly.  The calculation of the 3-body cluster contributions from the 
static part of $F_{ij}$ are also exact.  However, the 
3-body contributions from spin-orbit correlations 
and forces, the n$\geq$4-body contributions and the difference between the variational 
and the ground state energies are estimated.  The empirical $E_0(\rho)$ 
assumes $\rho_0=0.16$ fm$^{-3}$, $E_0(\rho_0)=-16$ MeV and an incompressibility of 
240 MeV.  The difference between the calculated and the empirical values 
is likely to reduce when the more realistic Illinois $V_{ijk}$ \cite{PPWC01} is 
used in place of the Urbana-IX.  However, a part of this difference is due to 
the approximations in the calculation.  

\begin{figure}[htbp]
\includegraphics[height=15cm, angle=-90]{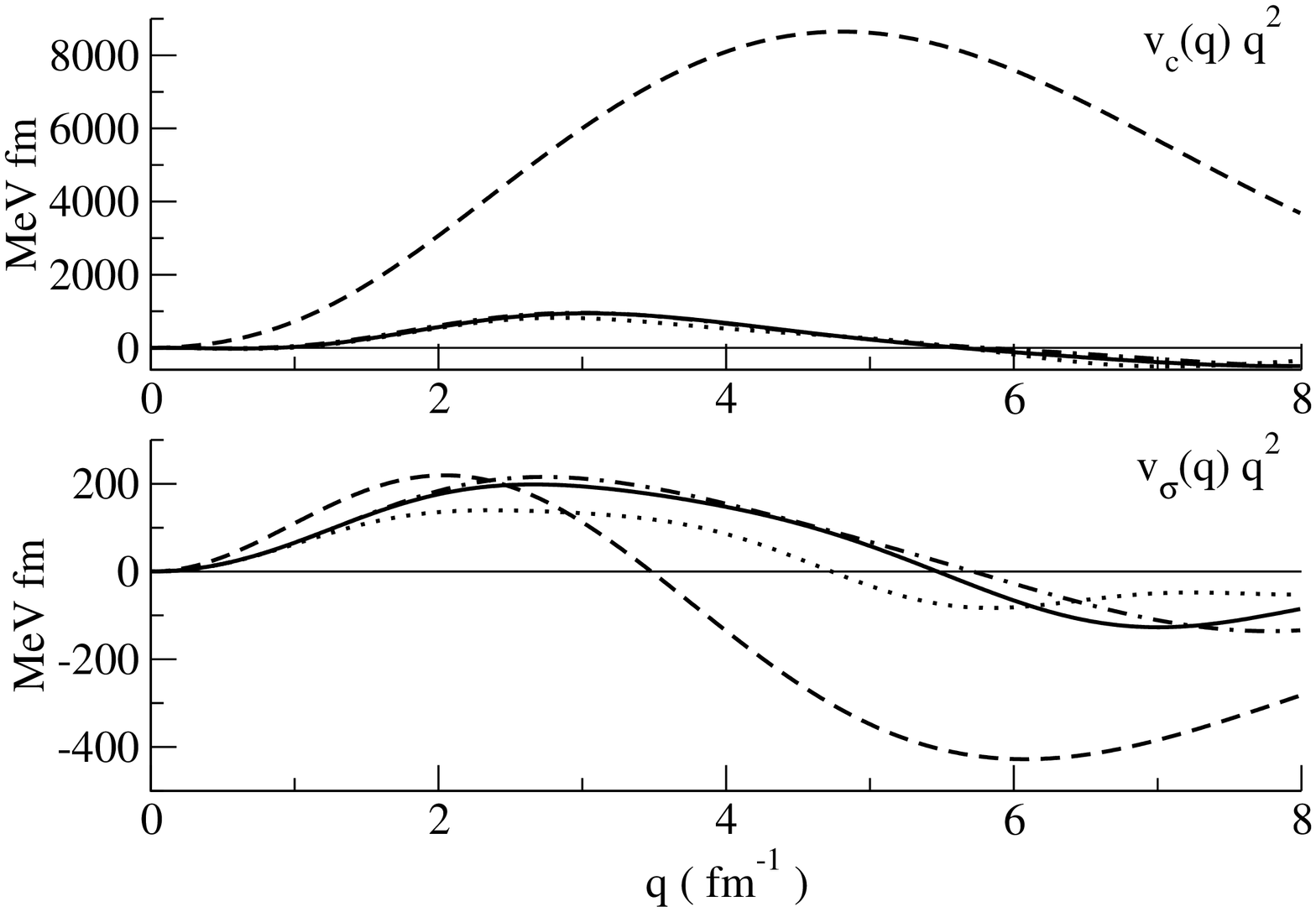}
\caption{The Fourier transform of the central and $\sigij$ components of $v^{CBS}$
 using $F_{ij}$ obtained at $\rho = \frac{1}{2}, 
1, \frac{3}{2} \rho_0$ are shown by dotted, solid, and dash-dot lines respectively.  The
dashed line shows the Fourier transform of the corresponding bare interaction.}
\label{cbi1}
\end{figure}

\begin{figure}[htbp]
\includegraphics[height=15cm, angle=-90]{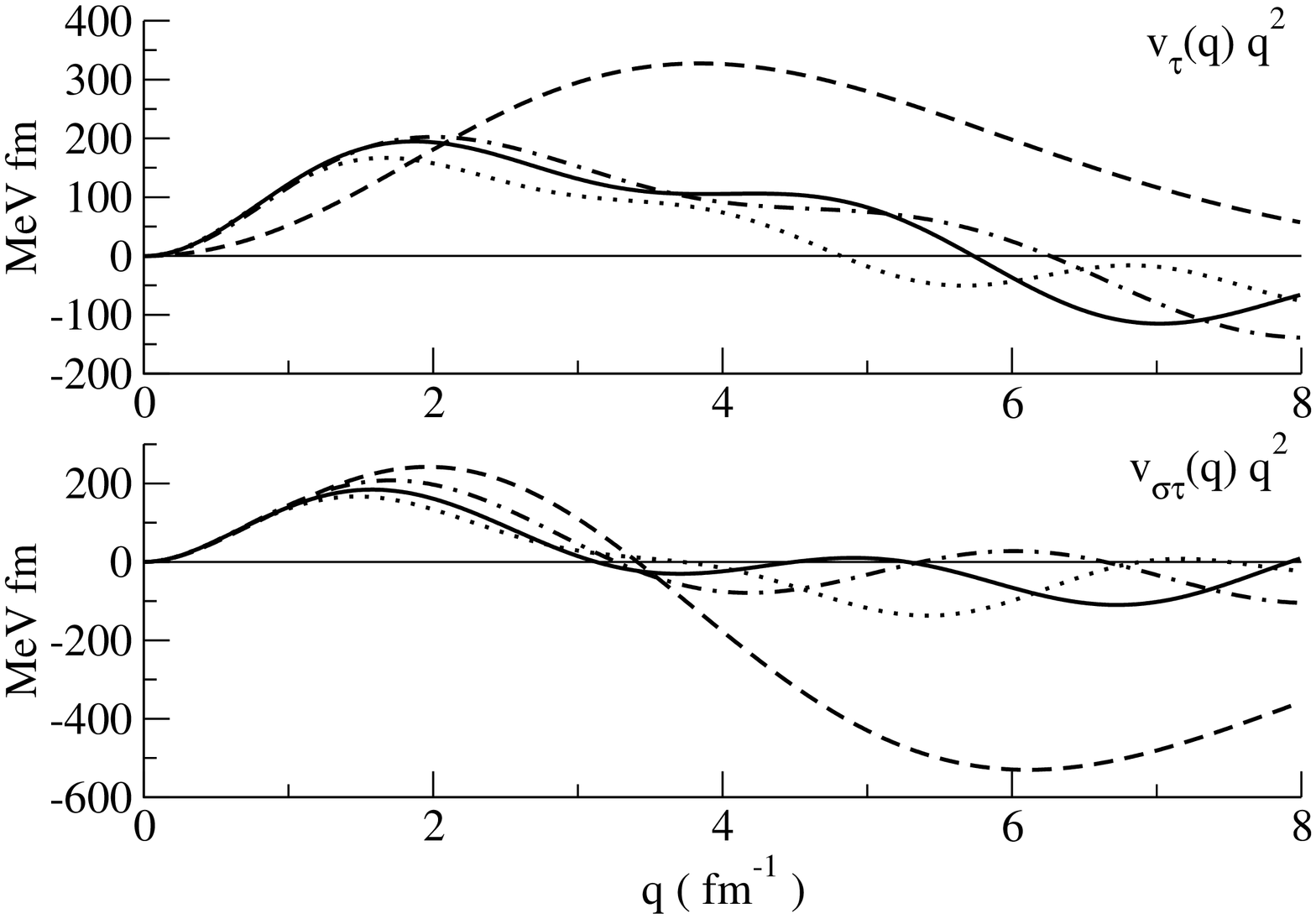}
\caption{The Fourier transform of the $\tauij$ and $\sigij \tauij$ components of
$v^{CBS}$ using $F_{ij}$ obtained at $\rho = \frac{1}{2},  1, \frac{3}{2} \rho_0$
are shown by dotted, solid, and dash-dot lines respectively.  
The dashed line shows the Fourier transform of the corresponding bare interaction.}
\label{cbi2}
\end{figure}

\begin{figure}[htbp]
\includegraphics[height=15cm, angle=-90]{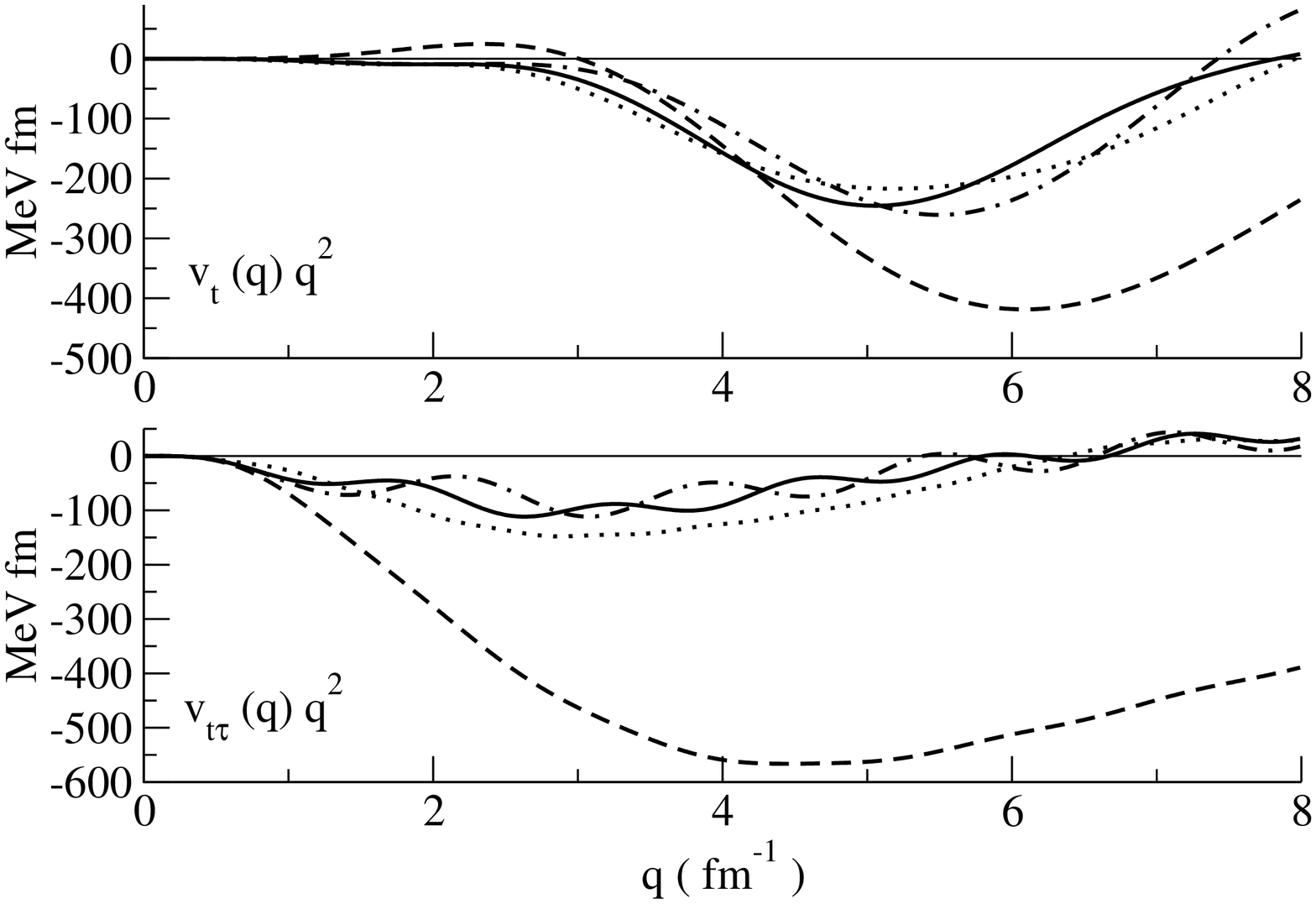}
\caption{The Fourier transform of the $S_{ij}$ and $\tauij S_{ij}$ components of
$v^{CBS}$ using $F_{ij}$ obtained at $\rho = \frac{1}{2},  1, \frac{3}{2} \rho_0$
are shown by dotted, solid, and dash-dot lines respectively.  
The dashed line shows the Fourier transform of the corresponding bare interaction.}
\label{cbi3}
\end{figure}

Next we consider the non-diagonal CBI.  Let a Fermi-gas state $|\Phi_F\rangle$ differ from 
$|\Phi_I\rangle$ in the occupation numbers of two single particle states:
\beq
|\Phi_F\rangle = a^{\dagger}_n a^{\dagger}_m a_j a_i |\Phi_I\rangle ~. 
\eeq
The matrix element of $H$ between the CB states is given by:
\beq
\langle F| H |I \rangle  
= \frac{\langle \Phi_F|[{\cal S} \Pi F_{ij}] \ H\ [{\cal S} \Pi F_{ij}]| \Phi_I \rangle}
{\sqrt{\langle \Phi_F| [{\cal S} \Pi F_{ij}]^2 | \Phi_F \rangle 
 \langle \Phi_I|[{\cal S} \Pi F_{ij}]^2| \Phi_I \rangle}}  ~.
\eeq
The numerator of this matrix element contains terms in which the kinetic energy 
operator acts on the $\Phi_I$.  They give:
\beq
\frac{\langle \Phi_F|[{\cal S} \Pi F_{ij}] [{\cal S} \Pi F_{ij}] T| \Phi_I \rangle}
{\sqrt{\langle \Phi_F| [{\cal S} \Pi F_{ij}]^2 | \Phi_F \rangle 
\langle \Phi_I|[{\cal S} \Pi F_{ij}]^2| \Phi_I \rangle}}  
=T_{FG}(I) \langle F |I \rangle  ~. 
\eeq
When the correlated states are orthogonalized this term is zero.  
Neglecting it the two-body 
cluster approximation of the above matrix element is obtained as:
\ba
\langle F| H |I \rangle &=& 
\langle mn|\Big[ v^{CBS} -\frac{1}{m}\Big(\boldnabla^{\prime}\cdot (F\boldnabla^{\prime})F 
+F (\boldnabla F)\cdot \boldnabla \Big) 
\Big]|ij \rangle  \nonumber \\
&=& \langle mn| v^{CB} | ij \rangle  ~. 
\ea
The $\boldnabla^{\prime}$ operate to the left while $\boldnabla$ to the right. 
When the momentum dependent term is negligible, this matrix is just the Fourier
transform  of $v^{CBS}$.  Using the algebra of operators $O^{p=1,6}_{12}$, 
and Eqs. (\ref{vcbs}) and (\ref{vcbsip}) we obtain: 
\beq
v_p^{CBS} = \sum_{q,r,s,t=1,6} f^qv^rf^s K^{qrt}K^{tsp} - \sum_{q,s=1,6} \frac{1}{m}  
f^q \Big( \nabla^2 - \frac{6}{r^2}(\delta_{s5}+\delta_{s6}) \Big) f^s K^{qsp}~. 
\eeq
Here we have used:
\beq
\nabla^2 f^t(r_{ij}) S_{ij} = S_{ij} \left( - \frac{6}{r^2_{ij}}f^t(r_{ij}) + 
\nabla^2 f^t(r_{ij}) \right) ~,
\eeq
and the $K^{pqr}$ matrices are given in ref \cite{PW79}.
The Fourier transforms of the $v^{CBS}_p$ are given in Figures \ref{cbi1} to \ref{cbi3}.  
Note that $S_{ij}=3\boldsigma_i {\hat{\bf q}}~\boldsigma_j {\hat{\bf q}} - \sigij $ 
in momentum space. 

The effective $v^{CBS}$ is weaker than the bare $v$, particularly at large values of 
$q$, as shown in Figs. \ref{cbi1}-\ref{cbi3}.  Perturbative corrections typically 
involve a loop integration over the momentum transfer ${\bf q}$ with a $q^2$ phase-space 
factor.  Hence in these figures we compare $q^2 v^{CBS}_p(q)$ with $q^2 v_p(q)$. 

\section{Conclusions}

We have calculated the effect of short range correlations on nuclear weak interaction 
matrix elements.  At low energies and small values of $q$ the charge current, 
weak transition rates 
are quenched by $\sim$ 20 to 25 \% in the simplest 2-body 
cluster approximation in 0th order CB theory.  This quenching is 
relatively independent of the density and 
proton fraction of nucleon matter as well as the momenta of nucleons in the
$\frac{1}{2}$ to $\frac{3}{2} \, \rho_0$ range.  However, it depends upon the momentum transfer $q$. 

The dominant part of the quenching is due to spin-isospin correlations induced by 
the OPEP in the bare interaction.  The OPEP changes the isospin of nucleons.  For 
example, in the $n \rightarrow p$ weak transition between uncorrelated states the 
active nucleon is initially a neutron and finally a proton with unit probability. In 
correlated states these probabilities are less than unit, and they reduce the weak 
interaction matrix elements.  In particular, for the Fermi case,  
most of the $q$ independent 
reduction is given by the product of the probabilities for the active 
quasi-nucleon to be initially a neutron and finally a proton.  A similar interpretation 
is also applicable for the GT matrix elements. 

In contrast to charge current, neutral current matix elements have a significant 
dependence on the proton fraction.  The neutron NV matrix element also depends on the 
total density, while the proton NV matrix element is very small and varies with all 
relevant parameters.  

We have also studied the effective nuclear interaction in the same CB used to calculate the 
weak interaction matrix elements.  The dominant static part of the lowest order 
2-body $v^{CB}$ gives 
fairly accurate results for the spin, isospin and spin-isospin susceptibilities 
of nucleon matter.  However, it is necessary to include at least 3-body effects 
to obtain the minimum in the $E_0(\rho)$ of symmetric nuclear matter.  The
$v^{CB}$ is much weaker than the bare $v$, and presummably can be used in perturbation 
theory formalism.

All calculations of weak transition rates using effective interactions must 
in principle use the quenched matrix elements calculated in the same basis. 
We plan to calculate the weak 
interaction rates in nucleon matter using the present effective 
operators and interactions.  To obtain more accurate predictions, it will be 
necessary to include $\geq$ 3-body 
terms in the cluster expansion of the CB effective operators and 
interactions. 

\begin{acknowledgments}
The authors would like to thank Drs. Chris Pethick, Sanjay Reddy and Jochen
Wambach for numerous discussions.  This work has been partially supported by the
US NSF via grant PHY 00-98353.
\end{acknowledgments}

\appendix

\section{Second Order Perturbation Theory} 

Standard perturbation theory is applicable when the bare interaction $v_{ij}$ is 
weak.  We then have $H=H_0+H_I$, $H_0=T$ and 
\beq
H_I = \sum_{i<j} v_{ij} ~, 
\eeq
Let $|\Phi_X \rangle $ be the unperturbed FG state. 
The perturbed, normalized state up to second order is given by:
\ba
|X \rangle &=& \Bigg( 1 - \frac{1}{2} \sum_{Y \neq X} 
	 \frac {|\langle \Phi_Y|H_I|\Phi_X \rangle |^2}{(E^0_{XY})^2} \Bigg) 
	 \Bigg( |\Phi_X \rangle + \sum_{Y \neq X} | \Phi_Y \rangle 
	 \frac {\langle \Phi_Y|H_I|\Phi_X \rangle }{E^0_{XY}} \nonumber \\
	 &+&  \sum_{Y,Z \neq X} | \Phi_Y \rangle 
	 \frac {\langle \Phi_Y|H_I|\Phi_Z \rangle }{E^0_{XY}} 
	 \frac {\langle \Phi_Z|H_I|\Phi_X \rangle }{E^0_{XZ}}
	 -  \sum_{Y \neq X} | \Phi_Y \rangle 
	 \frac {\langle \Phi_Y|H_I|\Phi_X \rangle }{E^0_{XY}} 
	 \frac {\langle \Phi_X|H_I|\Phi_X \rangle }{E^0_{XY}} \Bigg) ~,
\ea
$E^0_{XY}=T_{FG}(X)-T_{FG}(Y)$.  In this approximation the Fermi matrix element is 
given by $\langle F|O_F|I \rangle $, where $\Phi_I$ and $\Phi_F$ are
given by Eq. (\ref{phif}).

We are concerned only with two-body effects and
therefore consider only  the interactions $v_{aj}$ in $H_I$.  The last two
terms of the above $|X\rangle$  can be combined with the second by replacing
the $v_{aj}$ by an effective  interaction; hence we will omit them.  The direct
terms of $\langle F|O_F|I\rangle$ can be written as: 
\beq
\langle F|\sum_i O_F(i)|I\rangle_{direct} = F.0.d.a + F.1.d.j + F.2.d.j + F.2.d.a ~, 
\eeq
since $F.0.d.j$ and $F.1.d.a$ are zero.  $F.n.x.y$ is defined as in Section
\ref{CBFME} with the exception that $n$ here refers
to the order of $H_I$.  We obtain:
\ba
F.0.d.a &=& \langle {\bf k}_p| O_F(a) |{\bf k}_n \rangle = 1 \\ 
F.1.d.j &=& \sum_{{\bf h}_N} \langle {\bf k}_p,{\bf h}_N |O_F (j) \ \frac{Q}{E_0-H_0} \ v_{aj} |
{\bf k}_n , {\bf h}_N \rangle \nonumber \\
&+& \sum_{{\bf h}_N}  
\langle {\bf k}_p,{\bf h}_N | v_{aj} \ \frac{Q}{E_0+\omega-H_0} \ O_F (j) |
{\bf k}_n , {\bf h}_N \rangle \\
F.2.d.j &=& \sum_{{\bf h}_N} 
 \langle {\bf k}_p,{\bf h}_N |v_{aj} \ \frac{Q}{E_0+\omega-H_0} \
 O_F (j) \ \frac{Q}{E_0-H_0} \ v_{aj}|{\bf k}_n,{\bf h}_N \rangle  \\
F.2.d.a &=& \sum_{{\bf h}_N} \Bigg[
\langle {\bf k}_p,{\bf h}_N |v_{aj} \ \frac{Q}{E_0+\omega-H_0} \ O_F (a) \ 
\frac{Q}{E_0-H_0}  \ v_{aj}| {\bf k}_n,{\bf h}_N \rangle \nonumber \\ 
 &-& \frac{1}{2} 
\langle {\bf k}_p,{\bf h}_N | O_F (a) \ v_{aj} \ \frac{Q}{E_0-H_0} \ \frac{Q}{E_0-H_0} \ 
v_{aj}|{\bf k}_n,{\bf h}_N \rangle \nonumber \\
 &-& \frac{1}{2} 
\langle {\bf k}_p,{\bf h}_N | v_{aj} \ \frac{Q}{E_0+\omega-H_0} \ \frac{Q}{E_0+\omega-H_0} \ 
v_{aj} \ O_F (a) |{\bf k}_n,{\bf h}_N \rangle  \Bigg]
\ea
where $E_0= e({\bf k}_n) + e({\bf h}_N)$, $\omega = e({\bf k}_p)
- e({\bf k}_n)$, $Q$ is the projection operator to ensure Pauli exclusion in
intermediate states, and  ${\bf h}_{N}$ are any occupied proton or neutron
states.  We use $e({\bf k})$ to denote single particle energies;  
when $H_0 = T$, $e({\bf k}) = k^2/2m$.  

In order to make a connection with the correlated basis theory, we see that in
perturbation theory the unnormalized two-body wave function is given by:
\beq
\label{ptwf} |\Psi\rangle = \left( 1 + \sum_{i<j} \frac{Q}{E_0-H_0} v_{ij} \right) |\Phi \rangle~. 
\eeq
Comparing it with the correlated wave function (Eq. (\ref{cbwf})) we 
can identify: 
\beq
\label{fvrel}
(F_{ij}-1) \sim \frac{Q}{(E_0-H_0)} v_{ij} 
\eeq
when the interaction is weak.  In
reality, $v_{ij}$ is strong and Eq. (\ref{fvrel}) is not useful.  
The correlation operator is determined variationally and its $\omega$ 
dependence is neglected assuming that the average value of $E_0-H_0$ is 
much larger.  

It can be verified that all of the $F.n.d.y$ terms in 
Sect. II are obtained by replacing the: 
\beq
\frac{Q}{E_0-H_0} v_{aj} ~~ \mathrm{and} ~~ v_{aj} \frac{Q}{E_0+\omega-H_0} \nonumber 
\eeq
in Eqs. A4 to A7 by $(F_{aj}-1)$, since $F^{\dagger} = F$.

\section{The C- and F-coefficients}
\label{appen}

The C-parts required to calculate the effective weak 
vector operators in CB are obtained as follows:  Let $X,Y,Z$ be operators of type:
\beq
X=\sum_{p=1,6} x_p~O^p ~.
\eeq
The C-part of the product of operators is then given by:
\beq
C(XYZ) = \sum_{p,q=1,6} \sum_{r,s=1,6} x_p~y_q~z_r~K^{pqs}~K^{src} ~, 
\eeq
where $O^c \equiv 1$, and the $K^{pqr}$ are given in Ref. \cite{PW79}.  
The results are listed below. 
\ba
C^{11}_d &=& (\gfb)^2 + 3(\gfd)^2 + 6(\gff)^2 ~, \\
C^{01}_d = C^{10}_d &=& (\gfa-1)\gfb + 3 \gfc \gfd + 6 \gfe \gff  ~, \\
C^{00}_d &=& (\gfa-1)^2 + 3(\gfc)^2 + 6(\gfe)^2 ~, \\
C^{00}_e &=& (\gfa-1)^2 - 3 (\gfc)^2 + 12 (\gfe)^2 + 6 (\gfa-1) \gfc~, \\
C^{11}_e &=& (\gfb)^2 - 3 (\gfd)^2 + 12 (\gff)^2 + 6 \gfb \gfd ~,\\
C^{01}_e = C^{10}_e &=& (\gfa-1)\gfb - 3\gfc\gfd + 12 \gfe\gff + 3(\gfa-1)\gfd + 3\gfc\gfb~. 
\ea

The $\boldsigma_a$ and ${\bf A}_t$ parts of a product of $\boldsigma_a \cdot \boldsigma_j, 
~S_{aj}~,\boldsigma_a$ and $\boldsigma_j$ operators is 
obtained by repeated use of the Pauli identity:
\beq
\boldsigma \cdot {\bf B} \boldsigma \cdot {\bf C} = {\bf B}\cdot{\bf C} +
 i \boldsigma \cdot {\bf B} \times {\bf C} 
\eeq
to reduce it to terms linear in $\boldsigma_a$, $\boldsigma_j$.  Terms linear in
$\boldsigma_j$ go to zero on summing over $j$. The remaining 
terms linear in $\boldsigma_a$ are expressed in terms of the 
operators $\boldsigma_a$ and ${\bf A}_t$ to obtain the following equations. 

\ba
\gfooadc &=& (\gfa-1)^2 - (\gfc)^2 - 2 (\gfe)^2 ~, \\
\gfToadc=\gfoTadc &=& (\gfa-1) \gfb - \gfc\gfd - 2 \gfe \gff ~, \\
\gfTTadc &=& (\gfb)^2 - (\gfd)^2 - 2 (\gff)^2~, \\
\gfooadt &=& 4\gfc\gfe + 2(\gfe)^2~, \\
\gfToadt=\gfoTadt &=& 2\gfc\gff + 2\gfe\gfd + 2\gfe\gff~, \\
\gfTTadt &=& 4\gfd\gff + 2(\gff)^2~,
\ea
\ba
\gfoojdc &=& 2(\gfa-1)\gfc + 4 (\gfc)^2 - 4(\gfe)^2~, \\
\gfTojdc=\gfoTjdc &=& (\gfa-1)\gfd + \gfc\gfb + 2 \gfc\gfd - 2\gfe\gff~, \\
\gfTTjdc &=& 2 \gfb\gfd + 2 (\gfd)^2 - 2(\gff)^2~, \\
\gfoojdt &=& 2 (\gfa-1) \gfe - 2 \gfc\gfe + 2 (\gfe)^2 ~,\\
\gfTojdt=\gfoTjdt &=& (\gfa-1) \gff + \gfe\gfb - \gfc\gff - \gfe\gfd + 2\gfe\gff~, \\
\gfTTjdt &=& 2 \gfb \gff - 2 \gfd\gff + 2 (\gff)^2~. \\
\ea
\ba
\gfooaec &=& (\gfa-1)^2 + 2 (\gfa-1)\gfc  +  (\gfc)^2 -4 (\gfe)^2~, \\
\gfoTaec &=& (\gfa-1) \gfb - (\gfa-1) \gfd +3 \gfc \gfb +  \gfc \gfd  - 4 \gfe \gff~, \\
\gfToaec &=& 3 \gfd (\gfa-1) + \gfb (\gfa-1) + \gfd \gfc - \gfb \gfc - 4 \gff \gfe~, \\
\gfTTaec &=& 2 \gfd \gfb + (\gfb)^2 + (\gfd)^2 -4 (\gff)^2~, \\
\gfooaet &=& 2(\gfa-1) \gfe +2\gfc \gfe +  4 (\gfe)^2~, \\
\gfoTaet &=& 2(\gfa-1) \gff -2\gfc \gff + 4 \gfe \gfd + 4 \gfe \gff ~, \\
\gfToaet &=& -2\gfd \gfe + 4 \gff \gfc + 2\gfb \gfe + 4 \gff \gfe~, \\
\gfTTaet &=& 2\gfd \gff + 2\gfb \gff + 4 (\gff)^2~,
\ea
\ba
\gfoojec &=& (\gfa-1)^22\gfc +(\gfa-1)  - 4 (\gfe)^2 + (\gfc)^2 ~, \\
\gfoTjec &=& (\gfa-1) \gfb+ 3 (\gfa-1) \gfd-\gfc \gfb  - 4 \gfe \gff + \gfc \gfd ~, \\
\gfTojec &=& \gfd (\gfa-1) + \gfb (\gfa-1) - 4 \gff \gfe + \gfd \gfc + 3 \gfb \gfc~, \\
\gfTTjec &=& 2\gfd \gfb + (\gfb)^2 - 4 (\gff)^2 + (\gfd)^2 ~, \\
\gfoojet &=& 2\gfe (\gfa-1) +2\gfc \gfe +  4 (\gfe)^2~, \\
\gfoTjet &=& 4 \gfc \gff + 2\gfe \gfb - 2\gfe \gfd + 4 \gfe \gff~, \\
\gfTojet &=& 2\gff (\gfa-1)+4 \gfd \gfe  - 2\gff \gfc + 4 \gff \gfe~, \\
\gfTTjet &=& 2\gfd \gff + 2\gff \gfb + 4 (\gff)^2~.
\ea

\begin{table}[htbp]
\begin{tabular}{|c|c|c|c|}
\hline
$\,~ I~ \, $ & $\quad
P(a,\frac{1}{2} \rho_0) \quad$ & $\quad P(a,\rho_0) \quad$ & $\quad
P(a,\frac{3}{2} \rho_0) \quad$\\
\hline
$n \uparrow$ & 0.92 & 0.89 & 0.87 \\
$n \downarrow$ & 0.02 & 0.03 & 0.03 \\
$p \uparrow$&  0.02 & 0.03 & 0.03 \\
$p \downarrow$&  0.04 & 0.05 & 0.07 \\
\hline
\end{tabular}
\caption{Correlated Basis probabilities for the active quasi-nucleon $a$ to be
$N\uparrow$ and $N\downarrow$ in the initial state for $\rho =
\frac{1}{2},~1,~\frac{3}{2} \rho_0$ and $x_p=0.5$. The listed values include 
contributions of 1- and 2-body direct terms. } 
\label{table2}
\end{table}

\begin{table}[htbp]
\begin{tabular}{|c||c|c||c|c|}
\hline
$\quad \rho \quad $ & $\ |\langle O_F\rangle |^2(q=0)\ $ & 
$\qquad \alpha_F\qquad $ & $\ \gtav (q=0)\ $ 
& $\qquad \alpha_{GT}\qquad $ \\
\hline
0.08 & 0.80 &  $-$0.094  & 0.76 & 0.259\\
0.16 & 0.81 &  $-$0.075  & 0.75 & 0.060\\
0.24 & 0.86 &  $-$0.083  & 0.78 & 0.041\\
\hline
\end{tabular}
\caption{Quadratic fit to $\eta_F$ and $\gtav$ at small $q$.} 
\label{table1}
\end{table}

\begin{table}[htbp]
\begin{tabular}{|c||c|c|}
\hline
$\quad \rho \quad $ & $\quad C_p(\rho) \quad$ & $\quad C_n(\rho) \quad$\\
\hline
0.08 & 1.39 & $-$1.29\\
0.16 & 1.53 & $-$1.46\\
0.24 & 1.40 & $-$1.38\\
\hline
\end{tabular}
\caption{Linear fit to $\av$ for $x_p < 0.5$ at small $q$.}
\label{table3}
\end{table}

\begin{table}[htbp!]
\begin{tabular}{lrrr}
\hline
Density (fm$^{-3}$) &~~~~~0.08~&~~~~~0.16~&~~~~~0.24~\\
\hline
1-b $T_{FG}$ & 13.9 & 22.1 & 29.1 \\
2-b all      &$-$25.9 &$-$43.7 &$-$56.2 \\
3-b static   &    4.9 & 10.9 & 19.1 \\
\hline
3-b LS + $\geq$4-b all & $-$2.2 & $-$1.7 & 0.8 \\
$(E_0-E_V)$  & $-$0.6 & $-$1.8 & $-$3.3 \\
\hline
Calculated $E_0$ & $-$ 9.9 & $-$ 14.2 & $-$10.6 \\
Empirical~ $E_0$ & $-$ 12.1 & $-$ 16.0 & $-$12.9 \\
\hline
\end{tabular}
\caption{Contributions to the ground state energy of SNM from 
Argonne $v_{ij}$ and Urbana $V_{ijk}$ in MeV per nucleon} 
\label{t:snm}
\end{table}

\clearpage

\end{document}